\documentclass[a4paper,11pt]{article}

\usepackage[utf8]{inputenc}
\usepackage{mathtools,amsthm,amssymb}
\usepackage{tikz}
\usepackage{tabularray}
\usepackage{graphicx}
\usepackage[left=0.6667in, right=0.6667in, top=1in, bottom=1in]{geometry}
\usepackage[english]{babel}
\usepackage{csquotes}
\usepackage{titlecaps}
\usepackage[backend=biber, style=numeric-comp, sorting=nyt, sortcites, 
giveninits, maxbibnames=99]{biblatex}
\Addlcwords{a in of the for an or and}
\DeclareFieldFormat{titlecase}{\titlecap{#1}}

\usepackage{enumitem}
\usepackage{hyperref}

\graphicspath{{figures/}}
\DeclarePairedDelimiter{\abs}{\lvert}{\rvert}
\DeclarePairedDelimiter{\norm}{\lVert}{\rVert}
\DeclarePairedDelimiter{\ev}{\langle}{\rangle}
\DeclareMathOperator{\nimplies}{\ooalign{$\implies$\cr\cr$\mkern11mu\not$}}

\DeclareMathOperator*{\argmin}{arg\,min}
\DeclareMathOperator{\sgn}{sgn}

\newcommand*{\A}{\mathcal A}
\newcommand*{\R}{\mathbb R}
\newcommand*{\G}{\mathcal G}
\newcommand*{\B}{\mathcal B}
\newcommand*{\K}{\mathcal K}
\newcommand*{\U}{\mathcal U}
\newcommand*{\T}{\mathcal T}
\newcommand*{\Q}{\mathcal Q}
\newcommand*{\I}{I}

\newcommand*{\f}[1][]{g #1 (x,w,\tau)}
\newcommand*{\fxi}{\abs{\f}_\xi(w)}
\newcommand*{\signf}[1][]{\sgn_\xi #1 [\f](w)}
\newcommand*{\der}[2]{\frac{d#1}{d#2}}
\newcommand*{\whitebox}{\hbox{\hsize=0.75ex\vrule\vbox{\hrule\kern0.75ex\noindent\null
\hrule}\vrule}}
\newcommand*{\cut}{\delta_{\whitebox}}
\newcommand*{\dt}{\frac d{dt}}
\newcommand*{\phv}{\,\cdot\,}  % place-holder value

\newcommand*{\kNN}{$k$--$\mathit{NN}$\/}

\makeatletter
\newcommand*{\@intzstar}{\edef\p@wer{^2}\@basicintz}
\newcommand*{\@intznonstar}{\edef\p@wer{}\@basicintz}
\newcommand*{\@basicintz}{\enspace 
\mathclap{\int\limits_{\Omega\p@wer\times\I\p@wer}}\enspace}
\newcommand{\intz}{\@ifstar{\@intzstar}{\@intznonstar}}%

\makeatother

\let\phi=\varphi

\renewcommand*{\L}{A}
\renewcommand*{\H}{\mathcal H}
\renewcommand*{\(}{\begin{equation}}
\renewcommand*{\)}{\end{equation}}

\theoremstyle{plain}
\newtheorem{thm}{Theorem}
\newtheorem{prop}{Proposition}

\theoremstyle{remark}
\newtheorem{remark}{Remark}
\theoremstyle{definition}
\newtheorem{definition}{Definition}

\title{Breaking Consensus in Kinetic Opinion Formation Models on Graphons}
\author{Bertram D\"uring\thanks{Mathematics Institute, University of Warwick, 
Coventry, UK,  
\texttt{Bertram.During@warwick.ac.uk}} \and Jonathan 
Franceschi\thanks{Department of Mathematics, University of Pavia, Pavia, 
Italy, 
\texttt{jonathan.franceschi01@universitadipavia.it}} \and Marie-Therese 
Wolfram\thanks{Mathematics Institute, University of Warwick, Coventry, UK, 
\texttt{M.Wolfram@warwick.ac.uk}} 
\and Mattia Zanella\thanks{Department of Mathematics, University of Pavia, 
Pavia, Italy, \texttt{mattia.zanella@unipv.it}}}
\date{}

\definecolor{corr}{HTML}{cf1020}
\begin{document}
\maketitle
\begin{abstract}

In this work, we propose and investigate a strategy to prevent consensus in 
kinetic models for opinion formation. We consider a large interacting agent 
system, and assume that agent interactions are driven by compromise as well as 
self-thinking dynamics and also modulated by an underlying static social 
network. This network structure is included using so-called graphons, which 
modulate the interaction frequency in the 
corresponding kinetic formulation. We then derive the corresponding limiting 
Fokker--Planck equation, and analyze its large time behavior. This microscopic 
setting serves as a starting point for the proposed control strategy, which 
steers agents away from mean opinion and is characterized by a suitable 
penalization depending on the properties of the graphon. We show that this 
minimalist approach is very effective by analyzing the quasi-stationary mean-field model solutions in a plurality of graphon structures. Several 
numerical experiments are also provided to show the effectiveness of the 
approach in preventing the formation of  consensus steering the system toward 
a declustered state. 
\end{abstract}

\section{Introduction}

Opinion formation models have been extensively studied in several research 
communities in the last decades. Classical models are based on the assumption 
that an individual's opinion is influenced by binary interactions with others 
as well as their surroundings (for example through social media). Most of them 
describe the dynamics of each individual, resulting in complex large systems, 
see for example~\cite{degroot74, 
deffuant04,hegselmann02,galam08,iacomini23,motsch14,sznajd00, GGSP,jabin14}.
In many of these models the underlying microscopic dynamics lead to the 
formation of complex macroscopic patterns and collective states. Methodologies 
from statistical mechanics, especially kinetic theory, have been used 
successfully to derive and analyze these complex stationary states in suitable 
scaling limits. Toscani's seminal work on kinetic opinion formation models, 
see~\cite{toscani06}, was one of the starting points of various kinetic 
models for collective dynamics, studying for example the effects of leaders on 
the opinion formation process~\cite{APZ14,during09, during15}, decision 
making~\cite{pareschi17,pareschi19} and the influence of exogenous
factors~\cite{bondesan24, franceschi231,zanella_BMB23} to name a few.

There has been a general agreement that individuals change their opinion due to 
interactions with others (these interactions are almost always assumed to be 
binary). Most models assume that only like-minded individuals interact, known 
as bounded confidence models , and that dynamics are strongly driven by the 
tendency to compromise. In addition they often assume that individuals change 
their opinion due to self-thinking, for example because of exposure to 
different media channels. There is a rich literature on models for opinion 
formation in large interacting agent systems, see for 
example~\cite{hegselmann02,BL,motsch14} as well as the influence of social 
networks on 
the opinion formation process. The later trend was accelerated as more and more 
data from social networks, such as X (formerly known as Twitter) or Facebook 
became available. This allowed researchers to investigate for example 
challenging questions related to the influence of voter's behavior on the 
success of vaccination campaigns, see for example~\cite{albi23, albi17}. For 
other related works about the control of opinions on evolving networks 
see~\cite{APZ16}, polarization see~\cite{lee14,matakos17,amelkin17} and 
marketing aspects see~\cite{TTZ18}.

Several works proposing different control strategies to enhance consensus 
formation can be found in the multi-agent system literature, see for 
example~\cite{CFRT,CFTV}.
On the other hand control strategies to prevent consensus formation, we will 
refer these strategies as declustering, have been less 
studied~\cite{piccoli19}. However, these declustering strategies can be a 
useful tool to understand how for example software-managed social media 
accounts, also known as bots, can prevent consensus 
formation or steer opinions in social networks; see for 
example~\cite{alothali18,gilani19}. It is therefore of interest to understand 
control mechanisms, which prevent consensus formation in large social networks.

Social networks can be studied using graph theory. Graph theory has become one 
of the most active fields of
research in connection with the 
collective behavior of large populations of agents, see for 
example~\cite{barabasi99,barabasi09,barre17,barre18,newman03,watts98}. The 
necessity to 
handle millions, and often billions, of vertices lead to the study of 
large-scale statistical properties of graphs. In recent years, large discrete 
networks have been treated as continuous objects through the introduction of 
new mathematical structures called \emph{graphons}, which stands for \lq\lq 
graph functions\rq\rq, 
see~\cite{borgs14,lovasz12,cambridge-book,caron23,glasscock15}. The main 
feature of graphons relies on the possibility to bypass the introduction of 
classical adjacency matrix by looking at an associate function $W(x,y)$ 
encoding all the information on the connectivity of the original discrete graph.
Graph-theoretical research aside, the concept of graphons have been also 
applied to optimal control~\cite{gao19,hu23} and especially epidemiological 
theory~\cite{naldi22,dedios22,erol23}. We mention the recent works on kinetic 
and mean-field equations on 
graphons~\cite{bonnet22,bayraktar23,coppini22,nurisso23}: in 
particular, graphons are leveraged to analyze the behavior of mathematical 
models acting on networks when their number of nodes becomes very large. 
Thorough analysis of such limiting procedures have been also investigated 
recently---see e.g.,~\cite{medvedev14} and~\cite{petit21} and references 
therein---, increasing the interest on connections between differential 
equations models as continuous limits of stochastic processes happening on 
discrete graphs, like random walks and diffusion processes, for example.

Fokker--Planck type equations acting on suitable limits of 
large dense graphs in particular have been receiving increasing attention 
(\cite{coppini22, bayraktar23}). Within this framework, our analysis is devoted 
to a continuous model acting on a very large network (not necessarily dense) 
via its graphon representation rather than its limiting sequence of finite 
graphs. This allows for the very natural interpretation, from the kinetic 
theoretical point of view, of the network as a (continuous) interaction kernel, 
which then leads to an effective surrogate mean-field model that 
incorporates the graphon.

In this paper we propose and investigate possible strategies to prevent 
consensus formation in a kinetic model for opinion formation on networks. Our 
main contributions are 
\begin{itemize}
\item Development and analysis of a minimal control problem on the agent based 
level. 
\item Derivation of a closed form solution of the controlled model in suitable 
scaling limits, showing that the proposed strategy does indeed prevent 
consensus in the long time limit (under certain conditions on the parameters).
\item Provide extensive computational experiments illustrating the 
effectiveness of the proposed control strategy for power-law, \kNN\ and 
small-world graphons.
\end{itemize}

This paper is structured as follows: in Section~\ref{sec:general-model} we 
introduce and analyze a 
kinetic opinion formation model on a stationary network. In 
Section~\ref{sec:control} we 
propose a simple but very effective control mechanisms, which prevents the 
formation of consensus. In particular we are able to provide closed form 
optimal controls, 
which prevent consensus formation in certain parameter regimes. We corroborate 
our analytical results with computational experiments in 
Section~\ref{sec:numerics}.

\section{Kinetic models for opinion dynamics on graphon 
structures}\label{sec:general-model}

We consider a large population of indistinguishable agents each characterized 
by their opinion~$w$ belonging to~$\I \coloneqq  [-1, 1]$, where  $\pm 1$ 
corresponds to 
two opposite beliefs. Agents change their opinion through binary interactions 
with an interaction frequency modulated by an underlying static network. The 
opinion formation process itself is based on two different mechanisms:
\begin{itemize}
\item first compromise dynamics---so individuals with close opinion try to find 
a compromise 
\item and second opinion fluctuation, which are included via random variables.
\end{itemize} 
The interaction frequency of agents depends on the underlying graph structure, 
which we model by a graphon in this paper.

Before discussing the binary agent interactions, we recall some basics about 
graphons (and refer the interested reader to Appendix~\ref{sec:graphons} 
for a more detailed introduction as well as further references). Graphons are 
continuous objects that generalize the concept of simple graphs 
with a large number of vertices. In case of discrete graphs, nodes are usually 
referred to using the index $i = 1, 2, 
\ldots, N$, where $N$ is the number of vertices. However, in graphons the 
discrete set 
$\{1,\ldots, N\}$ is mapped onto the continuous interval $[0, 1]$, so that 
nodes are labeled as $x \in \Omega \subseteq [0,1]$, where $\Omega$ is a 
suitable subset of the unit interval of $\R$. We will therefore consider agents 
which are not only characterized by their individual opinion $w$ on a topic, 
but also their static position on the graphon $x\in \Omega\subseteq[0,1]$. \\

We consider the following setup for binary interactions: given two interacting 
agents characterized by their opinion and position in the graphon, that is 
$(x,w),(y,w_*) 
\in \Omega \times \I$ we compute their post interaction opinions 
$(x,w^\prime),(y,w_*^\prime)$ (note that their positions $x$ and $y$ in the 
graphon did not change) as: 
\(\label{eq:interaction}
\begin{aligned}
w' &= w - \gamma P(x,y)(w - w_*) + \eta D(x,w)\\
w_*' &= w_* - \gamma P(y,x)(w_* - w) + \tilde\eta D(x,w_*), 
\end{aligned}
\)
where $\gamma\in (0,1)$ is the so-called compromise parameter. The interaction 
function $P(\phv,\phv) \in [0,1]$ depends on the graphon 
coordinates $x,y \in \Omega$ and may also depend on the opinion variables 
$w,w_* \in \I$.
For instance, in \cite{toscani06} the case $P = 
P(\abs{w})$ as a non-increasing function with respect to~$\abs{w}$ and such 
that $0 \le P(\abs{w}) \le 1$ is explored. A different choice, resembling the case of a 
bounded confidence model like Hegselmann--Krause's~\cite{hegselmann02} can be 
obtained setting~$P = P(\abs{w - w_*}$). Moreover, we remark that this 
particular choice has the advantage of ensuring the conservation of the average 
opinion but presents analytical difficulties arising from the presence of the 
absolute value. In the rest of the manuscript, we focus on the case
where the interaction function only depends on the agents' positions on the graphon. A possible choice could be
\[	
P(x,y) = \exp(-\alpha\, d_i(x)/d_i(y)), \quad \alpha > 0,
\]
where $d_i(z)$ is the in-degree of the node at coordinate $z\in\Omega$ as 
defined in Definition~\ref{def:degree} of Appendix~\ref{sec:graphons}. 
Hence interactions depend on the connectivity of each agent---the more 
connected an agent is the less it is influenced by the other, while agents with 
a lower connectivity are affected more. In particular
\setbox0=\hbox{\fbox{$d_i(x) / d_i(y) \gg 1$}}
\begin{itemize}[left=1.1\wd0]
\item[$d_i(x) / d_i(y) \gg 1$] implies that, on average with respect to 
the random variable $\eta$, the agent with the highest degree keeps their 
opinion;
\item[$d_i(x) / d_i(y) \approx 1$] implies that agents with a similar number of 
incoming connections are the ones that can most influence each other;
\item[$d_i(x) / d_i(y) \ll 1$] implies that the less influential agents tend to 
adopt the opinion of the more connected ones.
\end{itemize}
A different possible choice of $P(x,y)$ that would give rise to similar 
dynamics would be, for instance, $P(x,y) = (1 + d_i(x)/d_i(y))^{-\alpha}$, with 
$\alpha > 0$.

In equation~\eqref{eq:interaction}, $\eta$ and $\tilde \eta$ are 
independent and identically distribution centered random variables
with finite moments up to order three and such that 
$\ev{\eta^2} =  \ev{\tilde\eta^2} = \sigma^2 < +\infty$. Here we denote by 
$\ev{\phv}$ the expectation with respect to the distribution of the random 
variables. The variables $\eta$, $\tilde\eta$ account for random fluctuations 
in an individual's opinion due to, for example, media exposure. The 
function $D(\phv, \phv)\ge0$ encodes the local relevance of the diffusion, 
possible choices include $D(w) = \sqrt{1-w^2}$. In this case agents diffuse the 
most if they have an indifferent opinion, that is $w \approx 0$, while they are 
less influenced by external factors once they settled on one of the two 
\lq extreme\rq\ choices. Note that this choice also ensures that opinions stay 
within $\I$.

Next we discuss some basic properties of the binary interaction defined before.
Under the assumption that the compromise propensity function $P$ satisfies $ 0 
\le P(x,y) \le 1$ and $0 < \gamma \le 1/2$, the following Proposition (see, 
e.g.~\cite{pareschi19,toscani06,TTZ18}) ensures that the post-interaction 
opinions still belong to the reference interval.
\begin{prop}
Assuming $0 \le P(x,y) \le 1$ and
\[
\left\lbrace
\begin{aligned}
\abs{\eta} &\le \ell,\\
\abs{\tilde\eta} &\le \ell,
\end{aligned}
\right.
\qquad \text{with } 
\ell \coloneqq \min_{w \in \I} \Bigl\lbrace \frac{(1 - w)}{D(w)}, D(w) 
\Bigr\rbrace,
\]
then the binary interaction~\eqref{eq:interaction} preserves the interval and 
the post-interaction opinions are such that $w^\prime$, $w_*^\prime \in \I$.  
\end{prop}%

A direct computation shows that for all $w$, $w_* \in \I$ 
\begin{equation}
\label{eq:micro_mean}
\ev{w' + w_*'} = w + w_* + \gamma(P(x,y)-P(y,x))(w_*-w). 
\end{equation}
If $P(x,y) = P(y,x)$, i.e., the compromise function $P$ is symmetric mean 
opinion is preserved in interactions, that is 
\[
\ev{w' + w_*'} = w + w_*.
\] 
On the other hand, the energy is not conserved on average since 
\[
\begin{split}
\ev{(w')^2 + (w_*')^2} &= w^2 + w_*^2 \\
	&\hphantom{{}=} + \gamma^2\bigl[ P^2(x,y)+P^2(y,x)\bigr](w_*-w)^2 + 
	2\gamma\bigl[ P(x,y)w-P(y,x)w_*\bigr](w_*-w) \\
	&\hphantom{{}=} + \sigma^2(D^2(x,w)+D^2(y,w_*)). 
\end{split}
\]
If $\sigma^2 \equiv 0$ and we have symmetric interactions, we see that the mean 
energy is dissipated.

We can now state the evolution equation for the distribution of agents  $f = f(x, w, t) $ with respect to their position $x \in \Omega$ and opinion $w \in \mathcal{I}$. Consider a fixed number of players, $N$, then the binary interactions \eqref{eq:interaction}
induce a discrete-time Markov process  with
$N$-particle joint probability distribution
$P_N(x_1,w_1,x_2,w_2,\dots,x_N, w_N,t)$.
This allows us to write a kinetic equation for the one-marginal distribution
function,
\[
P_1(x,w,t)=\int P_N(x, w,x_2,x_2,\dots, x_N ,x_N,t)\,
dx_2  dw_2 \cdots  dx_N dw_N,
\]
using only the one- and two-particle distribution functions \cite{Cerc88,CIP94},
\begin{multline*}
P_1(x, w,t+1)-P_1(x, w,t)=\\
\Bigg\langle \frac 1N \Biggl[\int
P_2(x_i,w_i, x_j, w_j,t) \bigl( \delta_0(x-x_i,w-w_i)+\delta_0(x-x_j,w-w_j) \bigr)\, dx_idw_i dx_jdw_j - 2P_1(x, w,t) \Biggr ]\Biggr\rangle.
\end{multline*}
Here, $\langle\cdot\rangle$ denotes the mean with respect to
the random variables $\eta,\tilde\eta$. By continuing this process one obtains a hierarchy of equations, the so-called
BBGKY-hierarchy \cite{Cerc88,CIP94}, describing the dynamics of the system of a large number of interacting agents.

It is a standard assumption to neglect correlations, implying that
\[
P_2(x_i,w_i, x_j, w_j,t)=P_1(x_i, w_i,t)P_1(x_j, w_j,t).
\]
By scaling time and performing the thermodynamical
limit $N\to\infty$, one  can use standard methods of kinetic theory
\cite{Cerc88,CIP94} to show that the time-evolution of the
one-agent distribution function $f$ is governed by  the
following non-Maxwellian Boltzmann equation
\(\label{eq:general}
\partial_t f(x,w,t) = Q(f,f)(x,w,t),
\)
where $Q(f,f)$ is the so-called collisional operator
\begin{equation}
\label{eq:coll}
Q(f,f)(x,w,t) =  \ev*{\enspace \intz \B(x,y)\Bigl(\frac{1}{{}'J} 
f(x,{}'w,t)\,f(y,{}'w_*,t) - f(x,w,t)f(y,w_*,t)\Bigr)\, dy \, dw_*}.
\end{equation}
Here $(x,{}' w)$ and $(y,{}' w_*)$ are pre-interaction opinions 
generating the post-interaction opinions  $(x, w)$ and $(y, w_*)$ and ${}' J$ 
is the Jacobian of the transformation $({}' w,{}' w_*)\to (w,w_*)$. In 
equation~\eqref{eq:coll} the kernel $\B(x,y) \colon \Omega^2 \to \R^+$ is a 
given graphon. It can be thought of as the continuous equivalent of an 
adjacency matrix. Its use in~\eqref{eq:coll} allows us to include an 
underlying network structure on the continuous level. Recent 
approaches to opinion formation modeling in the kinetic communities, 
e.g.,~\cite{albi23,APZ16,albi17,TTZ18}, take into account a 
graph structure via some of 
its statistical descriptions, like for example considering the number of 
connections of each agents as an adjoint variable. Instead, the use of a 
graphon kernel allows for a richer and more general description of individual 
connections among agents.

Solutions to equation \eqref{eq:general} preserve features of the underlying 
microscopic interaction rule. To compute the evolution of the mean opinion, we 
consider the weak formulation of equation~\eqref{eq:general}. Let 
$\varphi(x,w)$ be a test 
function, then
\begin{multline}
\label{eq:weak_coll}
\frac{d}{dt} \intz\varphi(x,w)f(x,w,t)\,dx\,dw  \\
=\frac{1}{2} \intz*\ev*{ 
\varphi(x,w') + \varphi(y,w_*') -\varphi(x,w) - \varphi(y,w_*)} 
f(x,w,t)f(y,w_*,t)dx\,dy\,dw\,dw_*. 
\end{multline}
Setting $\varphi(x,w)\equiv1$ in \eqref{eq:weak_coll} yields conservation of 
mass. Furthermore, from the conservation of the microscopic average opinion 
\eqref{eq:micro_mean} we see that $\varphi(x,w) = w$ gives conservation of the 
mean opinion (again assuming that the interaction function is symmetric).

Using conservation of the mean opinion, we can show that for 
any $\varphi(x,w) = w^\alpha \phi(x)$, where $\phi(\phv)$ is a suitable test 
function, the macroscopic quantities 
\begin{equation}
\label{eq:MA}
M_\alpha = \intz \phi(x) w^\alpha f(x,w,t)dx\,dw, 
\qquad 
\alpha = 0,1
\end{equation}
are conserved in time. Indeed, from~\eqref{eq:weak_coll} choosing 
$\varphi(x,w) = \phi(x) w^\alpha$ we get
\[
\begin{split}
\frac{d}{dt}M_\alpha &=\frac{1}{2} \intz 
\B(x,y)\left[\phi(x) \left\langle (w')^\alpha-w^\alpha\right\rangle + \phi(y) 
\left\langle 
(w_*')^\alpha-w_*^\alpha\right\rangle\right]f(x,w,t)f(y,w_*,t)dx\,dy\,dw\,dw_*. 
\end{split}
\]
From the above equation we get---for $\alpha = 0,1$---conservation of any 
weighted macroscopic moment of order $\alpha$ defined in \eqref{eq:MA}.

To investigate the second order moment of $f(x,w,t)$ we introduce the 
quantities
\(\label{eq:marginal-moments}
\Lambda(x,t) \coloneqq \int_\I w f(x,w,t)\, dw, \quad \text{ and } \quad 
\Xi(x,t) \coloneqq 
\int_\I w^2 f(x,w,t)\, dw, 
\)
that is, the first- and second-order moment, respectively, with respect to $w$ 
of agents with label~$x\in \Omega$. Clearly, integrating $\Lambda(x,t)$ and 
$\Xi(x,t)$ over $\Omega$ gives us the mean opinion and the energy of the 
population. \\
We will see that for bounded 
graphons $\B(x,y)$ and no diffusion, that is $\sigma=0$, the agent distribution 
$f(\phv,w,t)$ converges toward a Dirac delta 
distribution centered in the initial mean opinion. 
This is not surprising 
since the opinion dynamics corresponds to a consensus formation process 
modulated by the graphon. 
Indeed, we have for all $x \in \Omega$,
\[
\begin{aligned}
\dt \Xi(x,t) &=  
\int_\I\int_\I\int_\Omega  \B(x,y) \ev{({w'})^2 - w^2 } 
f(x,w,t)f(y,w_*,t)\, dy\, dw\, dw_*\\
&\le \norm{\B(x,y)}_{L^\infty(\Omega^2)} \int_\I\int_\I\int_\Omega
\bigl[2\gamma^2 P^2(x,y)(w-w_*)^2-2\gamma wP(x,y)(w-w_*)\bigr]\\
&\hphantom{{}= \norm{\B(x,y)}_{L^\infty(\Omega^2)} \int_\I\int_\I\int_\Omega} 
f(x,w,t) f(y,w_*,t) \, dy\, dw\, dw_*.
\end{aligned}
\]
Therefore, in the uniform interaction case $P(x,y)\equiv 1$, even in the 
presence of non-homogeneous graphon structure supposing $\norm{\B(x,y) 
}_{L^\infty(\Omega\times\Omega)}>0$, we get
\[
\dt \Xi(x,t) \le -2\gamma(1-\gamma)\norm{\B(x,y)}_{L^\infty(\Omega^2)} 
\bigl(\Xi(x,t) - \Lambda^2(x,t)\bigr).
\]
Thus, we have that the second order moment of  $f(\phv,w,t)$ tends to its mean 
squared, and its variance vanishes exponentially fast. Integrating both sides 
with respect to $x \in \Omega$ gives the estimate
\[
\dt E(t) \le -2\gamma(1-\gamma)\norm{\B(x,y)}_{L^\infty(\Omega^2)} \bigl(E(t) - 
m^2\bigr),
\]
where
\[
E(t) \coloneqq \int_{\Omega\times\I} w^2 f(x,w,t)\, dx\, dw,
\qquad m \coloneqq \int_{\Omega\times\I} w f(x,w,t)\, dx\, dw
\]
are the second order moment for the whole population and the global mean 
opinion, respectively. The latter quantity, thanks to 
equation~\eqref{eq:MA} is conserved in time. Therefore, a bounded graphon 
kernel implies exponential convergence of the distribution $f(x,w,t)$ toward a 
Dirac's Delta centered at the initial mean opinion.

\subsection{Derivation of a mean-field description}\label{sec:mean-field}

Since large time statistical properties of the introduced kinetic model are 
very 
difficult to obtain, several reduced complexity models have been proposed. In 
this direction, a deeper insight on the large time distribution of the 
introduced kinetic model can be obtained in the quasi-invariant regime 
presented in~\cite{toscani06}. The idea is to rescale both the interaction 
and diffusion parameters making the binary scheme~\eqref{eq:interaction} 
quasi-invariant. The idea has its roots in the so-called grazing collision 
limit of the classical Boltzmann equation, see~\cite{during09,pareschi13} and 
the references therein. The resulting model has the form of an 
aggregation-diffusion Fokker--Planck-type equation which is capable of 
encapsulating the information of microscopic dynamics and for which the study of asymptotic properties is typically easier.

We consider $\epsilon\ll 1$ and introduce the following scaling
\[
\gamma \mapsto \epsilon\gamma, \qquad \sigma \mapsto \sqrt\epsilon\sigma,
\]
for which $w'\approx w$ and $w_*' \approx w_*$. Next we Taylor-expand the term 
encoding the binary interactions in the weak form of the collision operator of 
equation~\eqref{eq:weak_coll}.
\[
\varphi(x,w') - \varphi(x,w) = \partial_w \varphi(x,w)(w'-w) + 
\dfrac{1}{2}\partial_w^2 \varphi(x,w) (w'-w)^2 + 
\dfrac{1}{6}\partial_w^3\varphi(w,\tilde w)(w'-w)^3,
\]
with $\tilde w \in (\min\{w,w'\},\max\{w,w'\})$. Introducing the new time 
variable $\tau = \epsilon t$ and the corresponding rescaled density 
$g(x,w,\tau) = f(x,w,\tau/\epsilon)$ we can rewrite \eqref{eq:weak_coll} as
\[
\begin{split}
\frac{d}{d\tau}\intz \varphi(x,w)g(x,w,\tau)\,dx\,dw 
&= \frac{1}{\epsilon} \intz* \B(x,y)\partial_w \varphi(x,w)\ev{w'-w} 
g(x,w,\tau)g(y,w_*,\tau)\,dx\,dy\,dw\,dw_* \\
&\hphantom{{}=} + \frac{1}{2\epsilon} \intz* \B(x,y)\partial_w^2 
\varphi(x,w)\ev{(w'-w)^2} g(x,w,\tau)g(y,w_*,\tau)\,dx\,dy\,dw\,dw_* \\
&\hphantom{{}=} + \frac{1}{\epsilon}R_{\varphi}(g,g),
\end{split}
\]
where 
\(\label{eq:remainder}
R_{\varphi}(g,g) = \frac{1}{6\epsilon} \intz \B(x,y) 
\partial_w^3 \varphi(x,w) \ev{(w'-w)^3} 
g(x,w,\tau)g(y,w_*,\tau)\,dx\,dy\,dw\,dw_*
\)
and $R_\varphi(f,f)/\epsilon \to 0$ under the hypothesis $\ev{\abs{\eta}^3} 
<+\infty$, see \cite{CPT,toscani06}. Consequently, in 
the limit $\epsilon \to 0^+$ we get
\begin{multline*}
\frac{d}{d\tau} \intz \varphi(x,w)g(x,w,\tau)\, dx\, dw \\
= \intz \B(x,y) \Bigl[
\gamma P(x,y)(w_*-w)\partial_w \varphi(x,w)+ \frac{\sigma^2}2 
D^2(x,w)\partial_w^2\varphi(x,w)\Bigr]g(x,w,\tau)g(y,w_*,\tau)\,dw_*\,dw\,dx\,dy
\end{multline*}
Therefore, integrating back by parts, we formally obtained a Fokker--Planck 
equation 
for the evolution of the distribution $g(x,w,\tau)$ 
\(\label{eq:general-FP}
\partial_\tau g(x,w,\tau) = \gamma\partial_w \bigl[ \K[g](x,w,t)g(x,w,\tau) 
\bigr] + 
\frac{\sigma^2}{2}\partial_{w}^2 \bigl[ \H[g](x,t) D^2(x,w) g(x,w,\tau)\bigr],
\)
where 
\(\label{eq:def_operators}
\begin{split}
\K[g](x,w,\tau)= & \intz  \B(x,y)P(x,y)(w-w_*) 
g(y,w_*,\tau)\, dy\, dw_*\\
\H[g](x,t)= &\intz \B(x,y) g(y,w_*,t)\, dy\, dw_*.
\end{split}
\)
The operator $\K$ corresponds to the network-modulated compromise process, the 
operator $\H$ corresponds to the network-weighted density $g(x,w,\tau)$. 
Equation~\eqref{eq:general-FP} is complemented with no-flux boundary 
conditions for all $x \in \Omega$:
\(\label{eq:general-bc}
\begin{aligned}
    \gamma  \K[g](x,w,t)g(x,w,\tau)  +\frac{\sigma^2}2 \partial_w 
    \bigl[\H[g](x,t) D^2(x,w) g(x,w,\tau)\bigr]\Big|_{w=\pm 1}&= 0,\\
    \H[g](x,t) D^2(x,w) g(x,w,\tau)\Big|_{w=\pm1} &= 0.
\end{aligned}
\)
This choice of boundary conditions ensures that 
system~\eqref{eq:general-FP}--\eqref{eq:general-bc} shares the same 
conservation properties as its microscopic kinetic counterpart. Indeed, we see 
that the mean opinion is conserved since
\(\label{eq:mean-preserved}
\frac{d}{dt} \intz wg(x,w,\tau)dw\,dx = \!\intz w \partial_w 
\Bigl[\gamma\K[g]\,g(x,w,\tau) + 
\frac{\sigma^2}{2}\H[g]\partial_w\bigl(D^2(x,w)g(x,w,\tau)\bigr) \Bigr] dw\,dx 
= 0,
\)
where we dropped for clarity the dependence on $x$, $w$ and $t$ for the 
operators $\K[g]$ and $\H[g]$. Furthermore, any macroscopic quantity of the 
form 
\(\label{eq:MFP}
M_\alpha^{\mathit{FP}} = \intz \phi(x)w^\alpha g(x,w,\tau)\, dw \,dx,
\)
is conserved in time. 

\subsection{Large time agent distribution}

Due to the presence of a general compromise propensity  
$P(\phv,\phv)$, a closed solution to equation~\eqref{eq:general-FP} 
is difficult to obtain. Nevertheless, under suitable assumptions on the 
graphon structure and on the diffusion function, we can write down a closed 
formulation for the large time agent distribution \eqref{eq:general-FP}.

In the following we restrict our analysis to the simplified 
situation where the interactions are homogeneous, i.e., $P(x,y)\equiv 1$, and 
the diffusion function is defined as
\(\label{eq:D}
D(x,w) = \sqrt{1-w^2} \qquad \text{for all $x \in \Omega$, }
\)
We recall that this choice of diffusion function ensures that $w$ stays within 
the domain $\I$. Furthermore, we suppose separability of the graphon $\B(x,y)$, 
which corresponds to 
\(\label{eq:separable-graphon}
\B(x,y) = \B_1(x)\B_2(y). 
\)
From the modeling point of view this choice is coherent with relevant examples 
of graphon structures, like the graphon associated to the case of scale free 
networks as proposed in~\cite{borgs14}. Indeed, in this case we have
\[
\B(x,y) = (xy)^{-\alpha}, \qquad 0 < \alpha < 1
\] 
satisfying the introduced separability assumption. Network structures that are 
found commonly in life and social sciences are often modeled using scale-free 
networks~\cite{cambridge-book, 
barabasi09, barabasi99}, i.e., simple graphs whose degree distribution 
possesses fat tails.

From \eqref{eq:MFP} we define the weighted mass and momentum as
\(\label{eq:rho-e-mu}
\rho \coloneqq \intz \B_2(y) g(y, w, \tau)\, dy\, dw,\qquad
\mu  \coloneqq \intz \B_2(y) w\,g(y, w, \tau)\, dy\, dw.
\)
Note that both quantities $\rho$ and $\mu$ are conserved in time.

Assuming relation~\eqref{eq:separable-graphon} holds, the steady state 
$g^\infty(x,w)$ of the Fokker--Planck model \eqref{eq:general-FP} satisfies the 
following equation
\begin{multline*}
\gamma \B_1(x)\biggl(\intz \B_2(y)(w-w_*)g^\infty(y,w_*)dw_*\,dy\biggr)  
g^\infty(x,w) \\
+ \frac{\sigma^2}{2}\partial_w \biggl[\B_1(x)\biggl( \intz 
\B_2(y)g^\infty(y,w_*)dw_* \,dy\biggr) D^2(x,w)g^\infty(x,w)\biggr] = 0.
\end{multline*}
Due to mass conservation and definitions \eqref{eq:rho-e-mu} we can simplify it 
further and obtain 
\[
\gamma\B_1(x) (\rho w - \mu) + \dfrac{\sigma^2}{2}\partial_w \left[\B_1(x)\rho 
\right] = 0.
\]
For our particular choice of diffusion function, that is \eqref{eq:D}, we 
can compute the steady state of $g^\infty(x,w)$ explicitly, 
see~\cite{toscani06}. In particular, {setting $\lambda = 
\sigma^2/\gamma$, we get
\(\label{eq:graphon-steady}
g^\infty(x,w) = \frac{\Gamma(2/\lambda)2^{1 - 2/\lambda}}{\Gamma\Bigl(\frac{1 + 
\mu/\rho}{\lambda}\Bigr)\Gamma\Bigl(\frac{1 - 
\mu/\rho}{\lambda}\Bigr)d_i(x)\cdot C_{\B}}\B_1(x) 
(1 + w)^{\frac{1 + \mu/\rho}{\lambda} - 1} (1 - w)^{\frac{1 - \mu/\rho} 
{\lambda} - 1}.
\)
which, as a function of the opinion, is a Beta distribution, weighted by the 
in-degree~$d_i(x)$ at $x \in \Omega$ times a graphon-dependent 
constant~$C_{\B}$ which depends on the way the splitting $\B(x,y) = 
\B_1(x)\B_2(y)$ is obtained and such that the right-hand-side of 
equation~\eqref{eq:graphon-steady} has unitary mass.

Model parameters appearing in the steady state allow to get insights on the 
shape and other characteristics of the equilibrium opinion distribution: for 
instance, when $\mu = 0$, $g^\infty$ is an even function of~$w$, so that the 
population has a neutral opinion on average. On the other hand, the balance 
between the actions of compromise and self-thinking dynamics expressed by 
the parameter~$\lambda$ tells us that if the action of the self-thinking is 
much stronger than the compromise one, i.e., $\lambda \gg 1$, then the tendency of 
the population would be to polarize at the extremes, tending to a mixture of 
Dirac's deltas at the boundary points of~$\I$. We refer the interested reader 
to~\cite{toscani06} for an in-depth analysis of the roles of interactions 
parameters on the equilibrium distribution.

\subsection{Analytical properties}

We continue by discussing some analytical properties for solutions to 
\eqref{eq:general-FP}. First we show that \eqref{eq:general-FP} preserves the 
$L^1$ 
regularity. To this end, we may rewrite equation~\eqref{eq:general-FP} as
\(\label{eq:FP}
\partial_\tau g(x,w,\tau) = \gamma\rho_P(x,\tau) \partial_w \bigl[(w - 
\mu_P(x,\tau))g(x,w,\tau)\bigr] +\frac{\sigma^2}{2}\H[g](x,\tau) 
\partial_w^2\bigl[D^2(w)g(x,w,\tau)\bigr].
\)
Note that we will again consider a specific form of diffusion, that is
$ D(w)=\sqrt{1-w^2}$. Furthermore, we introduce the 
following quantities, dependent on the compromise propensity function
\[
\rho_P(x,\tau) \coloneqq \intz \B(x,y)P(x,y) g(y, w, \tau)\, dy\, dw,\quad
\mu_P(x,\tau)  \coloneqq \frac{1}{\rho_P(x,\tau)}\intz \B(x,y)P(x,y) w\,g(y, w, 
\tau)\, dy\, dw.
\]
Note that $\mu_P(\phv, \tau)$ is well-defined since we are 
considering $P$ to be positive almost-anywhere and the graphon $\B$ to be 
nonnegative, as well.
Next, we take, for a given parameter~$\xi$, a regularized non-decreasing 
approximation of the sign function~$\sgn_\xi(\phv)$, and then introduce 
the anti-derivative of~$\signf$ for every~$w \in 
\I$ as the function~$\fxi$, where we stress the dependence on these 
functions on the variable~$w$. Now, let us fix $x \in 
\Omega$, multiply each side of equation~\eqref{eq:FP} and integrate 
with respect to~$w$. This gives
\[
\begin{aligned}
\der{}{\tau} \int_\I \fxi\, dw
	&= 	\gamma\rho_P(x,\tau)\int_\I \signf \Bigl[\partial_w [(w 
	-\mu(x,\tau))\f]\Bigr]\, dw\\ 
	&\hphantom{{}=}+ \frac{\sigma^2}{2}\H[g](x,\tau) \int_\I \signf 
	\Bigl[\partial_{w}^2 \bigl(D^2(w) 
	\f\bigr) \Bigr]\, dw\\
	&= -\gamma\rho_P(x,\tau)\int_\I \signf[']\partial_w \f \f\times\\
	&\hphantom{{}=}\times \bigl[w - \mu(x,\tau) + 	\partial_w D^2(w)]\bigr]\, 
	dw\\
	&\hphantom{{}=} - \frac{\sigma^2}{2}\H[g](x,\tau) \int_\I \signf[']
	\bigl[ D(w) \partial_{w}\f \bigr]^2\, dw,
\end{aligned}
\]
where we used the boundary conditions~\eqref{eq:general-bc}. Since
\[
\signf[']\partial_w \f \f = \partial_w\bigl[\signf \f - \fxi\bigr],
\]
we can substitute this expression and obtain
\[
\begin{aligned}
\der{}{\tau} \int_\I \fxi\, dw
	&= -\gamma\rho_P(x,\tau)\int_\I \partial_w\bigl[\signf \f - 
	\fxi\bigr]\cdot\\
	&\hphantom{{}=}\cdot\bigl[w - \mu(x,\tau) + \partial_w D^2(w)]\bigr]\, dw\\
	&\hphantom{{}=} - \frac{\sigma^2}{2}\H[g](x,\tau) \int_\I \signf[']
	\bigl[ D(w) \partial_{w}\f \bigr]^2\, dw.
\end{aligned}
\]
Now, the first integrand on the right-hand side vanishes as~$\xi 
\to 0^+$ if we integrate by parts one more time (since by construction 
$\lim_{\xi \to 
0^+} \signf \f = \abs{\f}(w)$ for almost every~$w \in \I$). This leaves us 
with 
the second integrand, which is nonnegative since we chose a non-decreasing 
approximation of the sign function. Finally, since graphons are nonnegative by 
definition at all points in their domain, we conclude that
\[
\der{}{\tau} \norm{\f}_{L^1([-1,1])} = \lim_{\xi \to 0^+} \der{}{\tau} \int_\I 
\fxi\, dw \le 0
\]
for all $x \in \Omega$. This implies that an initial datum in $L^1(\I)$ 
would ensure that $\f \in L^1(\I)$ for all $\tau > 0$.
\begin{remark}
The weak contractivity of the $L^1$ norm with respect to the opinion also 
allows us to prove uniqueness of solutions to~\eqref{eq:FP}. The proof is 
based on contradiction; assume there exist two 
solutions $\f$ and $s(x,w,\tau)$ and evaluate the regularized modulus 
of their difference for each point~$x \in \Omega$. 
If we fix $x \in \Omega$, then due to linearity with respect to $w$ we have 
that $\f - 
s(x,w,\tau)$ is a solution to equation~\eqref{eq:FP}, too. Therefore
\[
\lim_{\xi \to 0^+}\der{}{\tau} \int_\I \abs{(g - s)(x,w,\tau)}_\xi\, dw \le 0,
\]
which implies that, at $x \in \Omega$, $g = s$ for almost all $w \in \I$ 
and $\tau > 0$ since by construction we have $g(x,w,0) = s(x,w,0)$ for all $x 
\in \Omega$ and $w \in \I$. The claim then follows since
$x\in \Omega$ was chosen arbitrarily.
\end{remark}
\begin{remark}
The weak contractivity of the $L^1$ norm with 
respect to~$w$ (i.e., the norm is not increasing in time) gives us as a 
corollary that the model~\eqref{eq:FP} is positivity-preserving. The claim 
follows noting that its solution $\f$ has a 
vanishing negative part if the initial datum is nonnegative. Indeed, we can 
express the negative part of~$\f$ via the regularization we introduced 
earlier, that is
\[
\f[^-_\xi] = \frac12 (\fxi - \f), \quad \text{for all } x \in \Omega
\]
This way, if we integrate with respect to $w$, we have
\[
2\der{}{\tau} \int_\I \f[^-_\xi]\, dw = \underbrace{\der{}{\tau}\int_\I 
\fxi\, 
dw}_{\le 0} {}+ 0, \quad \text{for all } x\in \Omega,
\]
thanks to the first boundary condition in~\eqref{eq:general-bc}. Then it holds
\[
\der{}{\tau} \int_\I \f[^-]\, dw = \lim_{\xi \to 0^+} \der{}{\tau} \int_\I 
\f[^-_\xi]\, dw \le 0,
\]
for all $x \in \Omega$, which implies that the nonnegativity of the initial 
datum is preserved by the model~\eqref{eq:FP}.
\end{remark}

We conclude this section by extending the previous regularity result---if the 
initial datum is in $L^{p}(\I)$, $p>1$ at $x\in \Omega$ then the 
solution $g \in L^p(\I)$ at $x$ for all $\tau > 0$. We 
will use both the positivity-preserving and the $L^1$ regularity of its 
solution in the following. 

We note that we only show $L^p$-regularity for all $p \ge 2$ since the result 
will also hold also for $p \in (1, 2)$ due to the boundedness of the 
interval~$\I$. The idea is to rewrite 
equation~\eqref{eq:FP} and impose the associated no-flux boundary 
conditions in order to estimate the time evolution of $\norm{ g}_{L^p}$ using 
integration by parts. The hypothesis on 
$p$ to be greater or equal than~$2$ is needed since we will need to take the 
derivative of $\f[^{p-1}]$ under the integral sign, but as stated above it is 
not restrictive.

We define the right hand side of~\eqref{eq:FP} as $\Q(g,g)(x,w,\tau)$ that is
\[
\Q(g,g)(x,w,\tau) \coloneqq \gamma\rho_P(x,\tau)\biggl[\partial_w\bigl([(1 - 
\sigma^2)w - \mu_P(x,\tau)]\f\bigr) + 
\frac{\H[g](x,\tau)\sigma^2}{2\gamma\rho_P(x,\tau)}\partial_w\Bigl((1 - 
w^2)\partial_w\f\Bigr)\biggr].
\]
Suppose that the following no-flux boundary 
conditions hold for all~$x \in \Omega$ and for all $\tau > 0$
\(\label{eq:Lp-bc}
\begin{aligned}
    \gamma\rho_P(x,\tau)\bigl[((1 - \sigma^2)w - 
    \mu_P(x,\tau))\f\bigr]\Big|_{w=\pm 1}
    &= 0,\\
    \frac{\H[g](x,\tau)\sigma^2}{2\gamma\rho_P(x,\tau)}\bigl[(1 - w^2) 
    \partial_w \f \bigr]\Big|_{w=\pm 1} &= 0.
\end{aligned}
\)
Now let us multiply each side of equation~\eqref{eq:general-FP} by 
$\f[^{p-1}]$ and integrate with respect to~$w$
\[
\begin{aligned}
\frac1p \der{}{\tau}\norm{\f}_{L^p([-1, 1])}^p
	&= \int_\I \Q(g,g)\f[^{p-1}]\, dw\\
	&= \underbrace{\gamma\rho_P(x,\tau)\int_\I\frac{\partial}{\partial 
	w}\bigl([(1 - 
	\sigma^2)w - \mu_P(x,\tau)]\f\bigr)\f[^{p-1}]\, dw}_{\coloneqq \T_1}\\
	&\hphantom{{}=} + 
	\underbrace{\frac{\H[g](x,\tau)\sigma^2}{2\gamma\rho_P(x,\tau)}\int_\I 
		\frac{\partial}{\partial  w}\Bigl((1 
		- w^2)\frac{\partial}{\partial w}\f\Bigr)\f[^{p-1}]\, dw}_{\coloneqq 
		\T_2}.
\end{aligned}
\]
Now the goal is to show that $\T_2$ is non-positive, so that it can be ignored 
in the estimate, and then focus on $\T_1$. Starting with $\T_2$, we integrate 
by parts and use the second boundary condition 
in~\eqref{eq:Lp-bc} to obtain
\[
\T_2 = -\frac{\H[g](x,\tau)\sigma^2(p-1)}{2\gamma\rho_P(x,\tau)}\int_\I (1 - 
w^2) (\partial_w \f)^2 \f[^{p - 2}]\, dw \le 0,
\]
since $w \in \I$, $\f$ is nonnegative and $\rho_P(x,\tau)$ is 
nonnegative at all $x \in \Omega$. Next we consider $\T_1$, and derive two 
different estimates for it. If we expand the derivative 
with respect to~$w$ under the integral sign, we have
\[
\T_1 =  \gamma\rho_P(x,\tau)(1 - \sigma^2)\int_\I \f[^p]\, dw
	 + \gamma\rho_P(x,\tau)\int_\I \bigl[[(1 - \sigma^2)w - \mu_P(x,t)] 
	 \partial_w\f \f[^{p-1}]\bigr]\, dw.
\]
On the other hand, we could as well integrate by parts and using the first 
boundary condition in~\eqref{eq:Lp-bc} and get
\[
\T_1 = -\gamma\rho_P(x,\tau)(p - 1)\int_\I [(1 - \sigma^2)w - \mu_P(x,t)] 
		\partial_w \f \f[^{p-1}]\, dw.
\]
If we now use the identity
\[
\T_1 = \frac{p-1}{p} \T_1 + \frac1p \T_1
\]
to replace $\T_1$ as the appropriate convex combination of the two equations we 
obtain:
\[
\T_1 = \gamma\rho_P(x,\tau)(1 - \sigma^2)\int_\I \f[^p]\, dw.
\]
Putting everything together we deduce that
\[
\der{}{\tau}\norm{\f}_{L^p(\I)}^p \le p\gamma\rho_P(x,\tau)(1 - 
\sigma^2)\norm{\f}_{L^p(\I)}^p
\]
for a given $x \in \Omega$. Then Gronwall's lemma implies that if the initial 
datum belongs to $L^p(\I)$ at $x \in \Omega$, then $\f \in L^p(\I)$ 
at $x$ for all $\tau > 0$.

\section{Declustering: preventing consensus via control 
strategies}\label{sec:control}

In this section we focus on control strategies to prevent consensus. In 
particular, the one driven by the compromise process. This objective is 
different to more common optimal control strategies, 
which would for example steer the average opinion to a given 
target~\cite{ACFK,albi15,APZ16}.
To this end, we propose an additional interaction to 
\emph{prevent} the formation of opinion clusters by enforcing a controlled 
interaction. In particular, we consider a convex combination of two 
updates weighted by the parameter~$\theta \in (0, 1)$ such that a fraction 
$1-\theta$ of the population follows an opinion transition of the 
type~\eqref{eq:interaction}, whereas a fraction of size $\theta$ follows
an opinion update given by a controlled interaction of the form
\(\label{eq:general-general-control}
w'' = w - \gamma u^* S(w),
\)
where $u^*$ is an agent-based control arising from the solution 
of a suitable optimization problem and $S(w) \ge 0$ a suitable selection 
function dependent on the opinion. In particular, the optimization problem 
focuses on the minimization of a suitable convex cost functional~$\mathcal J$ 
on the set~$\U$ of admissible controls, which in our case are those such 
that the post-interaction opinion $w''$ stays within the interval $\I$ and has 
the form
\(\label{eq:control-problem-general}
u^* = \argmin_{u \in \U}\, \mathcal J(w''_*, u).
\)
The quantity~$w''_*$ appearing on the right-hand-side 
of~\eqref{eq:control-problem-general} is a \emph{virtual update} which the 
functional~$\mathcal J$ would be subject to: in fact, the optimal control 
$u^*$ is the solution of the following 
optimization problem
\(\label{eq:control-problem}
\left\lbrace
\begin{aligned}
u^* &= \argmin_{u \in \U} \Bigl(\frac12 (w''_* - m)^2 + \frac\nu2 u^2 
\Bigr),\\
w''_* &= w +\gamma u^* S(w),
\end{aligned}
\right.
\)
where $m$ denotes the average opinion of the population on the network at time $t$ 
and $\nu > 0$ is a regularization parameter. Notice that the actual update~$w''$ 
and the virtual update~$w''_*$ have an opposite effect on~$w$. 
Solving the associated Lagrange-multiplier problem
\[
(w''_* - m)\frac{\partial w''_*}{\partial u^*} + \nu u^* = 0, 
\]
gives
\[
u^* = -\frac{\gamma S(w)}{\nu + \gamma^2 S^2(w)} (w - m)
\]
and so the resulting interaction is
\(\label{eq:final-general-control}
w'' = w + \frac{\gamma^2 S^2(w)}{\gamma^2 S^2(w)+ \nu} (w - m).
\)
In particular, in the rest of the paper we focus on the selection function
\(\label{eq:S}
S(w) \coloneqq \chi( w \in \L), \qquad \L\coloneqq \Bigl(-\frac{\gamma^2(1 + m) 
+ \nu}{2\gamma^2 + \nu}, \frac{\gamma^2(1 + m) + \nu}{2\gamma^2 + \nu}\Bigr) 
\subseteq \I.
\)
%\textcolor{blue}{Why did you choose an update parameter $\gamma$ in the first 
%place (for the interaction with the control).}
\begin{remark}
Thanks to the presence of the indicator function in~\eqref{eq:S}, we can 
verify that $w'' \in [-1, 1]$ and therefore that the 
controlled interaction is admissible.
\end{remark}
 
We recall that we balance the two types of interactions: at a rate $1-\theta$, 
agents update their opinion according to~\eqref{eq:interaction}, at the 
rate~$\theta$ they interact with the external 
control~\eqref{eq:final-general-control}. This yields a kinetic model whose 
right-hand side is convex combination of two non-Maxwellian operators
\(\label{eq:general-control-equation}
\partial_t f(x,w,t) = (1- \theta) Q(f,f)(x,w,t)  + \theta Q_u(f)(x,w,t).
\)
The role of the parameter~$\theta \in [0,1]$ is to model the frequency at 
which the different kinds of interaction take place: it can be thought as the 
percentage of automated users (e.g., bots programmed by a third party) on the 
network. The operator $Q(f,f)$ is the same introduced in 
equation~\eqref{eq:coll}; the operator $Q_u(f)(x,w,t)$, instead, encodes the 
controlled update of agents' opinions as prescribed by the elementary 
interaction~\eqref{eq:final-general-control} and is therefore given by 
\(\label{eq:control-weak}
\intz\varphi(x,w)Q_u(f)(x,w,t) dw\,dx= 
\intz\B(x,y) 
(\phi(x,w^{\prime\prime})-\varphi(x,w))f(x,w,t)dw\,dx,
\)
for any test function $\varphi(\phv,\phv)$.

\subsection{Mean-field limit of the controlled model}

In this section we explicitly show how the introduced control is capable of
breaking consensus on the mean-field level for 
suitable choice of the penalization. We proceed like we did in 
Section~\ref{sec:mean-field} to derive a more 
approachable mean-field limit of 
equation~\eqref{eq:general-control-equation}, using the same scaling
\(\label{eq:scaling}
\tau \to t/\epsilon, \qquad \gamma \to \gamma\epsilon, \qquad \sigma^2 
\to \sigma^2\epsilon, \qquad \nu \to \kappa\epsilon,
\)
for a certain $\kappa \in \R^+$. In this case, particular care is needed in 
treating the weak form~\eqref{eq:control-weak} due to the presence of the 
indicator function in the interaction~\eqref{eq:final-general-control}. 
Using Taylor expansion like we did in Section~\ref{sec:mean-field}, yields
\[
\begin{aligned}
\frac{d}{d\tau} \intz \phi(x,w)g(x,w,\tau)\, dx\, dw
	&= (1-\theta)\biggl[-\intz\phi'(x,w)g(x,w,\tau)\K[g](x,w,\tau)\, dx\, dw\\
	&\hphantom{=(1-\theta)\biggl[}+ \intz 
	\phi''(x,w)g(x,w,\tau)\H[g](x,\tau) + \frac{1}{\epsilon}R_\phi(g,g)\biggr]\\
	&\hphantom{{}=}+\frac{\theta}{\epsilon}\biggl[\ \underbrace{\intz 
	\B(x,y)\partial_w\phi(x,w)(w'' - w)g(x,w,\tau)\, dx\, dw}_{{}\coloneqq 
	\A[g](x,w,\tau)} + 
	R'_\phi(g,g)\biggr].
\end{aligned}
\]
Here, $\K[g]$, $\H[g]$ and $R_\phi(g,g)$ are the same operators as in 
equations~\eqref{eq:def_operators} and~\eqref{eq:remainder} in
Section~\ref{sec:mean-field}, while we denote
\[
R'_\phi(g,g) = \frac{1}{2} \intz \B(x,y) 
\partial_w^2 \phi(x,w) (w'' - w)^2 
g(x,w,\tau)\, dx\,dw,
\]
which can be shown to go to zero with computations analogous to the ones for 
$R_\phi(g,g)$. We continue with the term $\theta\A[g](x,w,\tau)/\epsilon$:
\[
\begin{aligned}
\frac{\theta}{\epsilon}\intz \B(x,y)\partial_w\phi(x,w)(w'' - w)g(x,w,\tau)\, 
dx\, dw
	&= \theta\!\intz \B(x,y)\partial_w\phi(x,w) \frac{\epsilon 
	\gamma^2 S^2(w)}{\epsilon\gamma^2 S^2(w) + \kappa}(w - m) 
	g(x,w,\tau)\, dx\, dw\\
	&= \theta\enspace\!\mathclap{\int\limits_{\Omega\times\L}}\enspace
	\B(x,y) \partial_w\phi(x,w)	\frac{\gamma^2 S^2(w)}{\epsilon\gamma^2 S^2(w) 
	+ \kappa}(w - m) g(x,w,\tau)\, dx\, dw\\
	&\to \theta\intz \B(x,y) 
	\partial_w\phi(x,w)\frac{\gamma^2}{\kappa} (w - m)g(x,w,\tau)\, dx\, dw,
\end{aligned}
\]
since in the limit $\epsilon \to 0^+$ we have
\[
\L = \Bigl(-\frac{\epsilon^2\gamma^2(1 + m) 
+ \epsilon\kappa}{2\epsilon^2\gamma^2 + \epsilon\kappa}, 
\frac{\epsilon^2\gamma^2(1 + m) + 
\epsilon\kappa}{2\epsilon^2\gamma^2 + \epsilon\kappa}\Bigr) \to (-1, 1),
\]
where we recall that $\kappa > 0$ and $\abs m \le 1$. Therefore,  we obtain the 
following Fokker--Planck equation
\(\label{eq:general-control-FP}
\partial_\tau g(x,w,\tau) = \gamma\partial_w \bigl[ 
\K_\theta^u[g](x,w,\tau)g(x,w,\tau) 
\bigr] + \frac{\sigma^2}{2}\partial_{w}^2 \bigl[ \H_\theta[g](x,\tau) D^2(x,w) 
g(x,w,\tau)\bigr],
\)
where this time we define
\begin{align}\label{eq:K-u-theta}
\K_\theta^u[g](x,w,\tau) &\coloneqq (1-\theta)\intz
\B(x,y)P(x,y)(w-w_*) g(y,w_*,\tau)\, dy\, dw_*  
-\frac{\theta\gamma}{\kappa}\int_{\Omega}\B(x,y)(w - m)\, dy\notag\\
&= (1 - \theta)\K[g](x,w,\tau) -\frac{\theta\gamma 
d_i(x)}{\kappa}(w-m).
\end{align}
Here $d_i(x)$ is the in-degree of $x$ as defined in 
Section~\ref{sec:graphons} and $\H_\theta[g] \coloneqq (1 - \theta)\H[g]$. 
The associated boundary conditions which are necessary to perform the 
integration by parts are
\(\label{eq:general-control-bc}
\begin{aligned}
    \gamma \bigl[ \K_\theta^u[g](x,w,\tau)g(x,w,\tau) \bigr] +\frac{\sigma^2}2 
    \partial_w 
    \bigl[\H_\theta[g](x,t) D^2(x,w) g(x,w,\tau)\bigr]\Big|_{w=-1}^{w=1} &= 0,\\
 	\H_\theta[g](x,t) D^2(x,w) g(x,w,\tau)\Big|_{w=-1}^{w=1} &= 0.
\end{aligned}
\)
\begin{remark}
The mean opinion of the population is preserved. Indeed, multiplying each side 
of equation~\eqref{eq:general-control-FP} and then integrating by parts we get
\[
\begin{aligned}
\frac{d}{d\tau} \intz wg(x,w,\tau)dx\,dw &= (1-\theta)\intz w \partial_w 
\Bigl[\gamma\K[g]\,g(x,w,\tau) + 
\frac{\sigma^2}{2}\H[g]\partial_w\bigl(D^2(x,w)g(x,w,\tau)\bigr) \Bigr] dx\,dw\\
&\hphantom{{}=} +\frac{\theta\gamma d_i(x)}{\kappa}\intz (w-m)g(x,w,\tau)\, 
dx\, dw = 0,
\end{aligned}
\]
thanks to equations~\eqref{eq:K-u-theta} and \eqref{eq:mean-preserved}.

The evolution of the second order moment is trickier, due to the presence of 
general graphon kernel and the compromise propensity function. In case of the 
specific diffusion function $D(\phv,\phv)$ \eqref{eq:D} we can simplify the 
expression and obtain
\begin{equation}\label{eq:controlled-variance}
\begin{split}
\frac{d}{d\tau} \int_\I w^2 g(x,w,\tau)\, dw
	&= (1-\theta)\intz w^2 \partial_w \Bigl[ \bigl(\gamma \rho_P(x,\tau)(w - 
	\mu_P(x,\tau))g(x,w,\tau)\bigr) \\
	&\quad+ \frac{\sigma^2}{2} 
	\H[g](x,\tau) \partial_w 
	\bigl((1-w^2)g(x,w,\tau) \bigr)\Bigr] \, dx\, dw \\
	&\hphantom{{}=} +\frac{\theta\gamma d_i(x)}{\kappa}\intz w^2 
	\partial_w\bigl[(w-m)g(x,w,\tau)\bigr]\, dx\, dw\\
	&= \biggl[ -2(1-\theta)\Bigl(\gamma\rho_P(x,\tau) + 
	\frac{\sigma^2}{2}\H[g](x,\tau)\Bigr) + 
	\frac{2\theta\gamma^2d_i(x)}{\kappa} \biggr]\Xi(x,\tau) \\
	&\hphantom{{}=} -\biggl[2(1-\theta)\gamma\rho_P(x,\tau)\mu_P(x,\tau)
 	+ \frac{2\theta\gamma^2 d_i(x)m}{\kappa} \biggr]\Lambda(x,\tau)  \\
	&\quad+ (1-\theta)\sigma^2\Phi(x,\tau)\H[g](x,\tau),
\end{split}
\end{equation}
where we note $\Phi(x,\tau) \coloneqq \int_\I g(x,w,\tau)\, dw$. 
Equation~\eqref{eq:controlled-variance} is still quite general: further 
insights on the trend of the second order moment can be found in the 
specialized setting of Remark~\ref{rem:energy}. 
\end{remark}

\subsection{Effects of the penalization coefficient on the quasi-equilibrium 
distribution}

In the following we will compute the quasi-equilibrium distribution 
of~\eqref{eq:general-control-FP}. This corresponds to solving
\[
\gamma\partial_w \bigl[ \K_\theta^u[g](x,w,\tau)g(x,w,\tau) \bigr] + 
\frac{\sigma^2}{2}\partial_{w}^2 \bigl[ \H_\theta[g](x,\tau) D^2(x,w) 
f(x,w,t)\bigr] = 0
\]
which we can write in closed form (due to the no-flux bc 
\eqref{eq:general-control-bc}). Then the quasi-equilibrium 
distribution~$f^{\rm qe}$ is given by
\[
g^{\rm qe}(x,w,\tau) = C\, \exp\left(-\int_{-1}^w \frac{\gamma 
\K_\theta^u[f](x,v,\tau) + 
\sigma^2/2 \, \H_\theta[g](x,\tau)\, \partial_v D^2(x,v)}{\sigma^2/2\, 
\H_\theta[g](x,\tau)\, D^2(x,v)}\, dv\right),
\]
where $C$ is a normalizing constant and under the formal assumption that 
$\H[g](x,\tau)D^2(x,v)$ can be taken to be nonzero almost everywhere at $x\in 
\Omega$ and on $[-1, 1]$.

To compare the controlled case to the one obtained in 
equation~\eqref{eq:graphon-steady} we focus on the case $D(x,w) = \sqrt{1 - 
w^2}$, and, having chosen a unitary compromise propensity function we can write
\(\label{eq:quasi-equilibrium}
g^{\rm qe}(x,w,\tau) = C (1 - w)^{\alpha_-} (1 + w)^{\alpha_+},
\)
where we define
\[
\begin{aligned}
\alpha_-(\theta,\gamma,\sigma,\kappa,\tau) &\coloneqq 
\frac{\gamma\bigl[\kappa(1-\theta)(\rho_p(x,\tau) - \mu_p(x,\tau)) - 
\gamma\theta d_i(x)(1 - m)\bigr]}{\kappa\rho_p(x,\tau)(1-\theta)\sigma^2} - 1,\\
\alpha_+(\theta,\gamma,\sigma,\kappa,\tau) &\coloneqq 
\frac{\gamma\bigl[\kappa(1-\theta)(\rho_p(x,\tau) + \mu_p(x,\tau)) - 
\gamma\theta d_i(x)(1 + m)\bigr]}{\kappa\rho_p(x,\tau)(1-\theta)\sigma^2} - 1.
\end{aligned}
\]
If we fix~$x \in \Omega$, the quasi-equilibrium state of 
equation~\eqref{eq:quasi-equilibrium} is a Beta distribution with respect to 
the variable~$w$. In fact, taking $\theta = 0$ and a separable graphon kernel 
$\B(x,y)$ in equation \eqref{eq:quasi-equilibrium} gives us precisely the 
steady state we found for the uncontrolled problem, given 
by~\eqref{eq:graphon-steady}.

We can exploit our knowledge of the quasi-equilibrium to influence the level of 
declustering of the system: indeed, the opinion distribution is in its least 
clustered form when it is uniform, i.e., when $\alpha_- = \alpha_+ = 0$. If we 
impose these constraints, we can solve them for the penalty term $\kappa$ as a 
function of the network position~$x\in \Omega$ and of time~$\tau$, i.e., 
$\kappa=\kappa(x,\tau)$. We obtain
\(\label{eq:kappa-bar-general}
\left\lbrace
\begin{aligned}
\kappa(x,\tau) &= \frac{\gamma^2\theta 
d_i(x)}{\rho_P(x,\tau)(1-\theta)(\gamma - 
\sigma^2)}\\
m &= 0\\
\mu_P(x,\tau) &= 0,
\end{aligned}
\right.
\)
which, under the further hypothesis that $\B(x,y) = \B_1(x)\B_2(y)$ and $P(x,y) 
\equiv 1$, simplifies to
\(\label{eq:kappa-bar}
\bar\kappa = \frac{\gamma^2\theta}{(1-\theta)(\gamma - \sigma^2)}.
\)
\begin{remark}\label{rem:kappa-bar}
The constraint $m = 0$ in \eqref{eq:kappa-bar-general} is necessary to have a 
uniform distribution over a symmetric domain, while imposing $\mu_1(x,\tau) = 
0$ is needed to have $\alpha_- =0$ and $\alpha_+ = 0$ simultaneously. 

When $g \approx U(\Omega\times[-1, 1])$, that is the distribution is almost 
uniform over the domain $ \Omega\times[-1, 1]$, we have $\rho_1(x,\tau) \approx 
d_i(x)$, so that
\[
g^{\rm qe}(x,w,t) \to g^\infty(x,w) \implies \kappa(x,\tau) \to 
\bar\kappa, \qquad \text{for all } x\in\Omega.
\]
Finally, we stress that the choices~\eqref{eq:kappa-bar-general} and 
\eqref{eq:kappa-bar} that would appear in the controlled 
update~\eqref{eq:general-general-control} come from the analysis of a 
quasi-equilibrium state for the distribution~$g(x,w,\tau)$ which has been 
computed from the Fokker--Planck equation~\eqref{eq:general-control-FP}. This 
implies that the penalty term would be effective in achieving the declustering 
effect only for a parameter regime in which $\epsilon$ is sufficiently small.
\end{remark}
\begin{remark}\label{rem:energy}
If we consider again the evolution of the second order 
moment as in~\eqref{eq:controlled-variance}, assume that  $P(\phv,\phv) \in 
L^\infty(\Omega\times \Omega)$, and use the 
$\kappa = 
\kappa(x,\tau)$ defined in 
equation~\eqref{eq:kappa-bar-general}, we obtain
\[
\frac{d}{d\tau} \Xi(x,\tau)\le 
	-3\norm{P(x,y)}_{L^\infty(\Omega^2)}(1-\theta)\H[g](x,\tau)\sigma^2 
	\Bigl[\Xi(x,\tau) - 
	\frac{\phi(x,\tau)}{3} + \bigl(\gamma\mu_P(x,\tau) + (\gamma - 
	\sigma^2)m\bigr)\Lambda(x,\tau)\Bigr].
\]
This estimate can be further simplified if $m = \Lambda(x,\tau) = 0$ for all 
$x\in \Omega$ (a condition needed to observe a centered uniform distribution):
\[
\frac{d}{d\tau} \Xi(x,\tau)
	\le -3\norm{P(x,y)}_{L^\infty(\Omega^2)}(1-\theta)\H[g](x,\tau)\sigma^2 
	\Bigl[\Xi(x,\tau) - 
	\frac{\phi(x,\tau)}{3}\Bigr].
\]
In particular, by Gronwall's inequality we have that whenever the graphon 
kernel is bounded, the variance converges exponentially in time toward~$1/3$, 
since, integrating both sides with respect to~$x\in\Omega$ gives
\(\label{eq:exponential-convergence-uniform}
\frac{d}{d\tau} E(\tau) \le 
-3\norm{P(x,y)}_{L^\infty(\Omega^2)}\norm{\B(x,y)}_{L^\infty(\Omega^2)}(1-\theta)\sigma^2
 (E(\tau) - 1/3),
\)
where $1/3$ is the variance of the uniform distribution over $\Omega\times\I$.
\end{remark}

\section{Numerical tests}\label{sec:numerics}

We conclude by illustrating the declustering strategy with various 
computational experiments. We show first the consistency of the 
quasi-invariant limit of the controlled 
model~\eqref{eq:general-control-equation} 
in the network-homogeneous case. In this case, we choose a uniform, constant 
graphon kernel $\B(x,y)\equiv c \in (0,1]$; in particular,  $\B(x,y) \equiv 1$.
In the second example we check the quasi-invariant limit using the power-law 
graphon, which is separable, to model 
an interaction happening on a scale-free network. In the third experiment we 
illustrate the dynamics for non-separable, graphon kernels.
All tests were performed using direct simulation Monte Carlo methods for the 
Boltzmann equation~\eqref{eq:general-control-equation}; we refer 
to~\cite{pareschi13,pareschi19} and references therein for further details.

\begin{table}
\centering
\begin{tblr}{column{1-2}={mode=dmath},
			 row{1}={mode=text},
			 column{1-2}={c},
			 column{3} = {l},
			 hline{1,Z} = {2pt, solid},
			 hline{2} = {1pt, solid}}
Parameter & Value(s) & Definition\\
N & 10^5 & Number of agents\\
\theta & 10^{-1} & Rate of interaction with the controlled 
update~\eqref{eq:final-general-control}\\
\epsilon & \{ 10^{-1},\, 10^{-2},\, 5\times 10^{-4} \}& Scaling parameter of 
equation~\eqref{eq:scaling}\\
\Delta t & \epsilon & Time discretization\\
T & \{ 8, 32 \} & Final time of the simulation\\
\gamma & \epsilon & Compromise parameter as in~\eqref{eq:coll}\\
\sigma^2 & \gamma/4 & Variance of random variables $\eta$, $\tilde\eta$ as 
in~\eqref{eq:coll}\\
\kappa(x,\tau) & & Penalization coefficient as 
in~\eqref{eq:kappa-bar-general}\\
\bar\kappa & 4/27 & Specialized 
penalization coefficient as computed 
in~\eqref{eq:kappa-bar}
\end{tblr}
\caption{List of parameters used within numerical experiments.}
\label{tab:par}
\end{table}

We start describing our method by rewriting 
equation~\eqref{eq:general-control-equation} in strong form as a 
sum of gain and loss parts:
\(
\begin{aligned}
\partial_t f(x,w,t) &= (1- \theta) Q(f,f)(x,w,t)  + \theta Q_u(f)(x,w,t)\\
	&=  (1-\theta)\intz \B(x,y)\Bigl(\frac{1}{{}'J} 
	f(x,{}'w,t)\,f(y,{}'w_*,t) - f(x,w,t)f(y,w_*,t)\Bigr)\, dy \, dw_*\\
	&\hphantom{{}=} + \theta \intz \B(x,y)\Bigl(\frac{1}{{}''J} 
	f(x,{}''w,t) - f(y,w_*,t)\Bigr)\, dy\,  dw_*.
\end{aligned}
\)
We indicate with $Q^\Sigma$ and $Q_u^\Sigma$ the operators obtained replacing 
the graphon kernel $\B(x,y)$ with the approximated version $\B^\Sigma(x,y$, 
given by
\[
\B^\Sigma(x,y) \coloneqq \min\{ \B(x,y), \Sigma \},
\]
where $\Sigma$ is an upper bound for $\B(x,y)$ over $\Omega^2$. Whenever we 
consider an unbounded interaction kernel (e.g., the power-law case) we consider 
a suitable truncation for $\Sigma$. If we now highlight the gain and loss parts 
of $Q^\Sigma$ and $Q_u^\Sigma$, we have
\[
\begin{aligned}
\partial_t f(x,w,t) &= (1-\theta)\biggl[Q^{\Sigma+}(f,f) + 
f(x,w,t)\ev[\bigg]{\intz 
\bigl(\Sigma - \B^\Sigma(x,y)\bigr)f(y,w_*,t)\, dy\, dw_*} -\Sigma 
f(x,w,t)\Bigr]\\
	&\hphantom{{}=} \theta\Bigl[Q_u^{\Sigma+}(f) + f(x,w,t)\ev[\bigg]{\intz 
	\bigl(\Sigma - \B^\Sigma(x,y)\bigr)f(y,w_*,t)\, dy\, dw_*} -\Sigma 
	f(x,w,t)\biggr],
\end{aligned}
\]
where we define
\[
\begin{aligned}
	Q^{\Sigma+} &\coloneqq \ev[\bigg]{\intz B^\Sigma(x,y) \Bigl(\frac{1}{{}'J} 
	f(x,{}'w,t)\,f(y,{}'w_*,t)\Bigr)\, dy, \, dw_*},\\
	Q_u^{\Sigma+}&\coloneqq  \ev[\bigg]{ \intz \B(x,y)\frac{1}{{}''J} 
	f(x,{}''w,t)\, dy\,  dw_*}.
\end{aligned}
\]
Then, we discretize the time interval $[0, T]$ with time step $\Delta t > 0$ 
and denote as $f^n(x,w)$ the time approximation $f(x,w,n\Delta t)$ to consider 
the forward-Euler-type scheme
\[
f^{n+1} = (1 - \Sigma\Delta t) f^n  + \Sigma\Delta t \frac{\mathcal S(f^n, 
f^n)}{\Sigma},
\]
where we define
\[
\begin{aligned}
\mathcal S(f,f) &\coloneqq (1-\theta)\biggl[Q^{\Sigma+}(f,f) + 
f(x,w,t)\ev[\bigg]{\intz 
\bigl(\Sigma - \B^\Sigma(x,y)\bigr)f(y,w_*,t)\, dy\, dw_*} -\Sigma 
f(x,w,t)\biggr]\\
	&\hphantom{{}=} \theta\biggl[Q_u^{\Sigma+}(f) + f(x,w,t)\ev[\bigg]{\intz 
	\bigl(\Sigma - \B^\Sigma(x,y)\bigr)f(y,w_*,t)\, dy\, dw_*} -\Sigma 
	f(x,w,t)\biggr].
\end{aligned}
\]
We remark that under the condition $\Sigma \Delta t \le 1$, $f^{n+1}$ is 
well-defined as a probability density. 

We report in Table~\ref{tab:par} all the parameters we used in our 
computational experiments. Moreover, we always fix $D(x,w) = \sqrt{1 - w^2}$ as 
diffusion function and $P(x,y) \equiv 1$ as compromise tendency function. 
Finally, we consider as initial 
distribution a state close to full consensus represented by a 
truncated Gaussian distribution over the interval $\mathcal I$ for all $x \in 
\Omega$
\[
f_0(w) = 
\begin{cases}
C \displaystyle \exp\left(\frac{(w-u_0)^2}{2\sigma_0^2}\right), & w \in [-1,1], \\
0 &\text{otherwise},
\end{cases}
\]
where $u_0 = 0$, $\sigma_0^2 = \frac{1}{10}$ and $C>0$ is a normalization constant.

\subsection{Consistency of the mean-field model: the network-homogeneous case}

We take model~\eqref{eq:general-control-FP} with $\B(x,y) \equiv 1$, which 
corresponds to a fully connected network, in which every node is adjacent to 
every other. Using the quasi-invariant scaling, we again approximate the 
dynamics of~\eqref{eq:general-control-equation} by the one-dimensional 
Fokker--Planck equation
\[
\partial_\tau g(w,\tau) = 
\gamma\partial_w\bigl[1 - \theta(1 + \gamma/\kappa)(w-m)g(w,\tau)\bigr]
 + (1-\theta)\frac{\sigma^2}{2}\partial_w^2\bigl[(1 - w^2)g(w,\tau)\bigr].
\]
Note that we choose $\kappa = \bar\kappa$, as in 
equation~\eqref{eq:kappa-bar}, which corresponds to the optimal scaling to 
ensure declustering. From Figure~\ref{fig:homogeneous-dist} we can see that 
as $\epsilon$ approaches zero, the controlled update gets fully effective and 
the state relaxes toward a uniform distribution.
\begin{figure}[htbp]
\centering
\hbox to \textwidth{%
\includegraphics[width=0.3\textwidth]{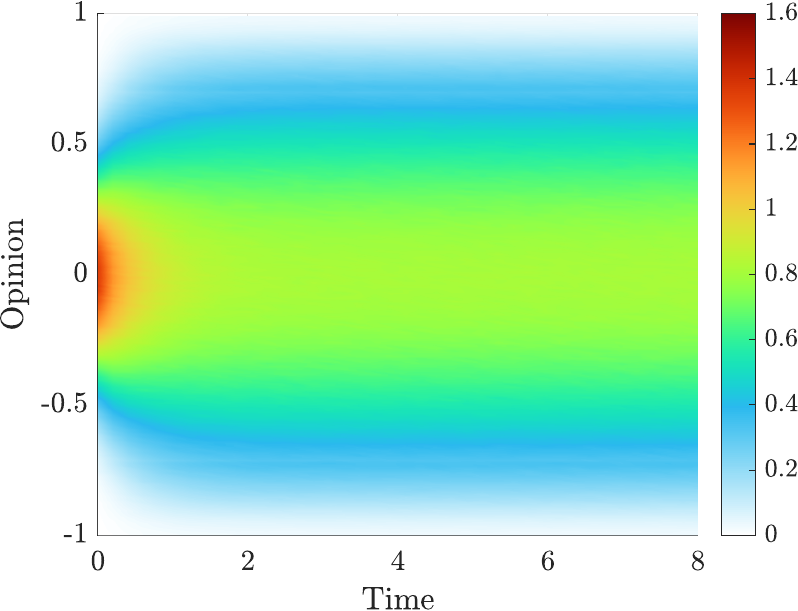}\hfil
\includegraphics[width=0.3\textwidth]{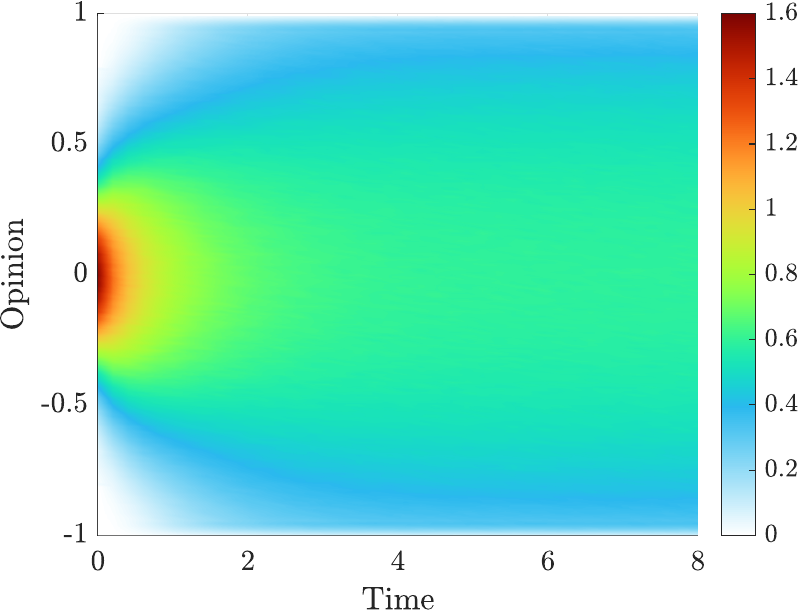}\hfil
\includegraphics[width=0.3\textwidth]{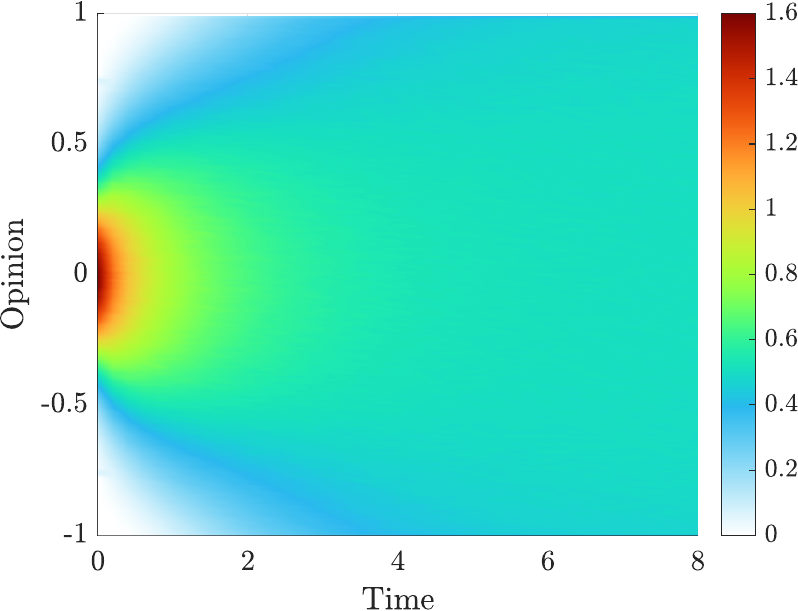}}
\caption{Time evolution of the opinion distribution $g(w,\tau)$ for the 
network-homogeneous case with different choices of parameter~$\epsilon$, 
respectively $\epsilon = 10^{-1}$, $\epsilon = 10^{-2}$ and $\epsilon = 5\times 
10^{-4}$ from left to right.}
\label{fig:homogeneous-dist}
\end{figure}
This is also testified by Figure~\ref{fig:homogeneous-steady-entropy}, where 
we report the profile of the distribution~$g(w,\tau)$ at time $\tau = T$, 
with $T = 8$. We also report the evolution from $\tau = 0$ to $\tau = T$ of the 
entropy, computed as
\[
H[g](\tau) = -\int_\I g(w,\tau)\log g(w,\tau)\, dw.
\]
We can see that when $\epsilon \ll 1$ the entropy approaches the 
value~$\log(2)$, which corresponds to the entropy of the uniform distribution 
over the interval~$\I$.
\begin{figure}[htbp]
\setbox0=\hbox{\includegraphics[width=0.45\textwidth]{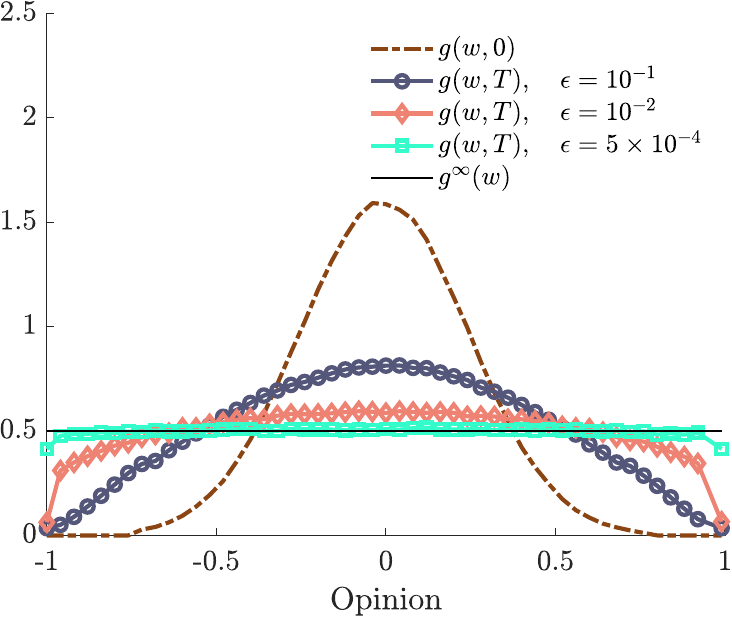}}
\centering
\hbox to \textwidth{%
\includegraphics[width=0.45\textwidth]{homogeneous-steady-new-eps-converted-to.pdf}\hfil
\includegraphics[width=0.45\textwidth,height=\ht0]{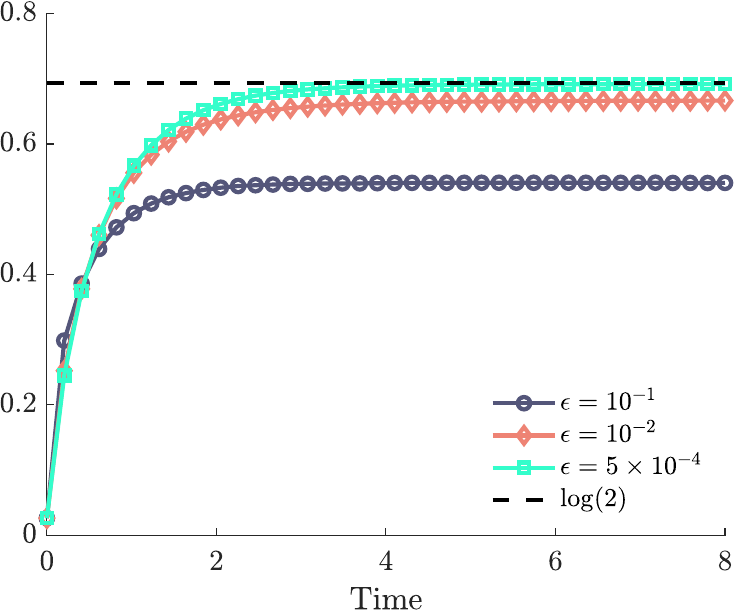}}
\caption{Left: comparison of the distribution $g(w,T)$ for various values of 
$\epsilon$, as in Figure~\ref{fig:homogeneous-dist}. Right: comparison of the 
time evolution for the entropy~$H[g](\tau)$ for the same values 
of~$\epsilon$.}
\label{fig:homogeneous-steady-entropy}
\end{figure}

\subsection{Consistency of the mean-field model: the power-law network case}
Next, we take model~\eqref{eq:general-control-FP} with $\B(x,y) =
9/16(xy)^{-1/4}$, i.e., the power-law graphon. We recall that this special 
choice yields equation~\eqref{eq:general-FP}. Since the power-law 
graphon is separable, the optimal value for the penalty term is $\bar\kappa$ of 
equation~\eqref{eq:kappa-bar} as for the network-homogeneous case.
Figure~\ref{fig:pl-evolution} shows the evolution of $g(x,w,\tau)$ for 
different values of~$\epsilon$. Since the distribution depends on both the 
opinion and the 
network position, we illustrate $g(x,w,\tau)$ at three instances in time in the 
first row, that is $\tau = 0$, $\tau = T/2$ and $\tau = T$, where this time we 
fix $T = 
32$. The second row of plots in Figure~\ref{fig:pl-evolution} shows the 
opinion marginal $\int_\Omega g(w,t)\, dx $ for different values of $\epsilon$.
\begin{figure}[htbp]
\centering
\hbox to \textwidth{%
\includegraphics[width=0.3\textwidth]{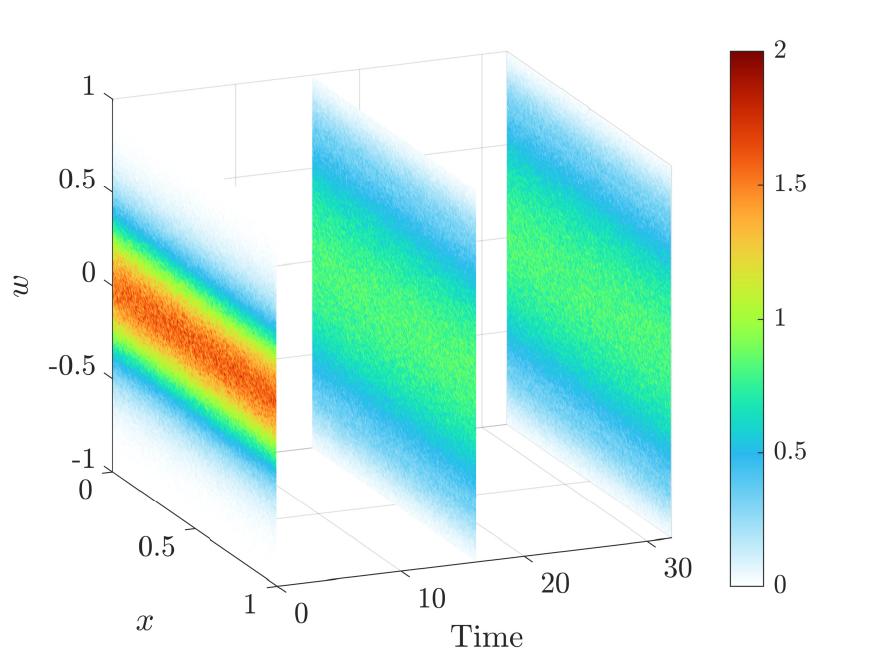}\hfil
\includegraphics[width=0.3\textwidth]{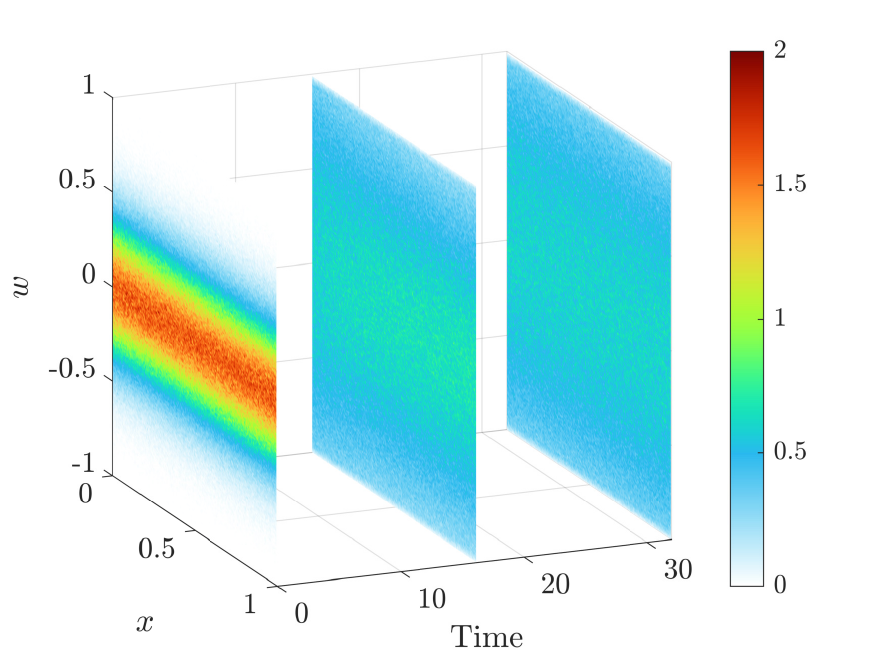}\hfil
\includegraphics[width=0.3\textwidth]{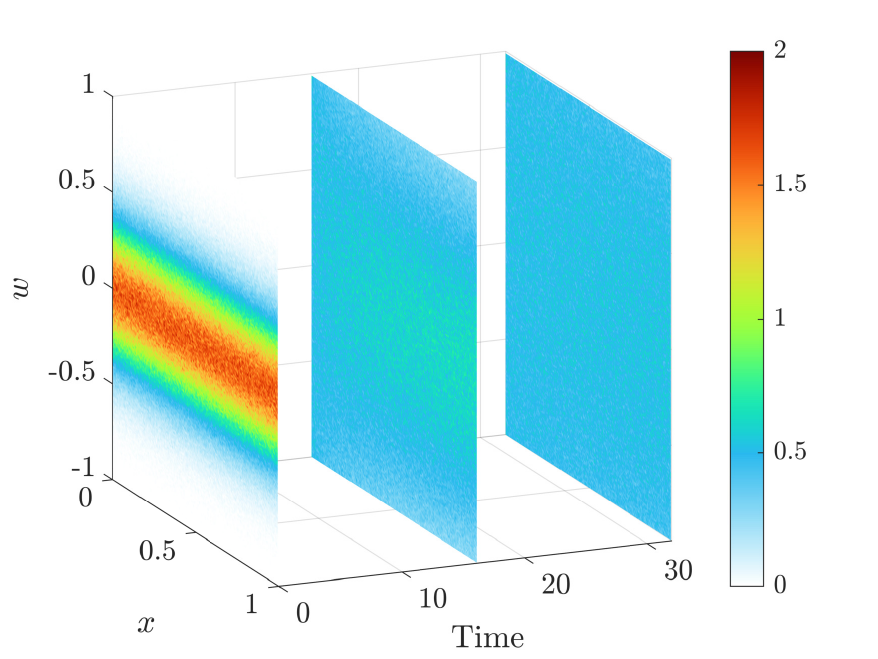}}
\hbox to \textwidth{%
\includegraphics[width=0.3\textwidth]{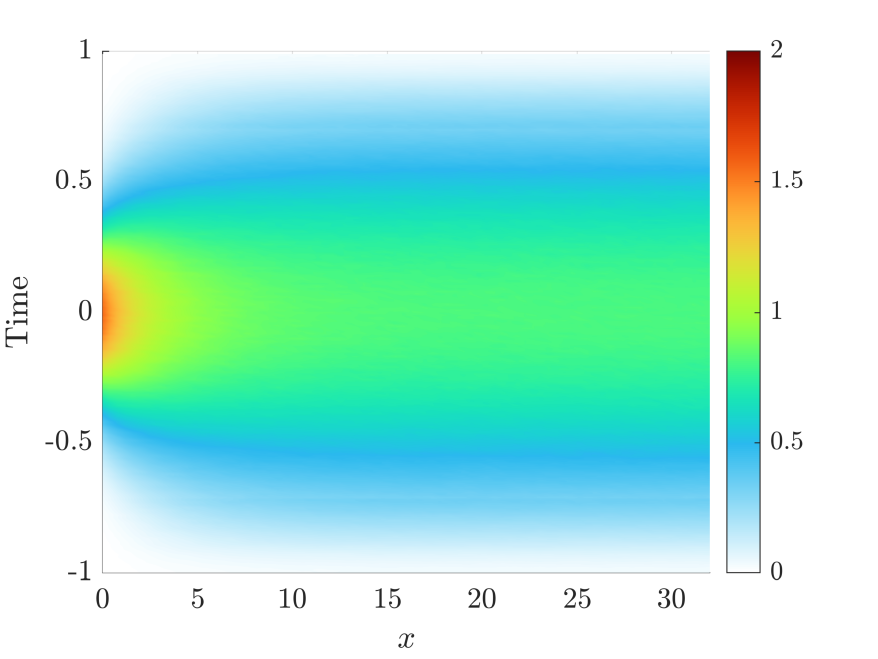}\hfil
\includegraphics[width=0.3\textwidth]{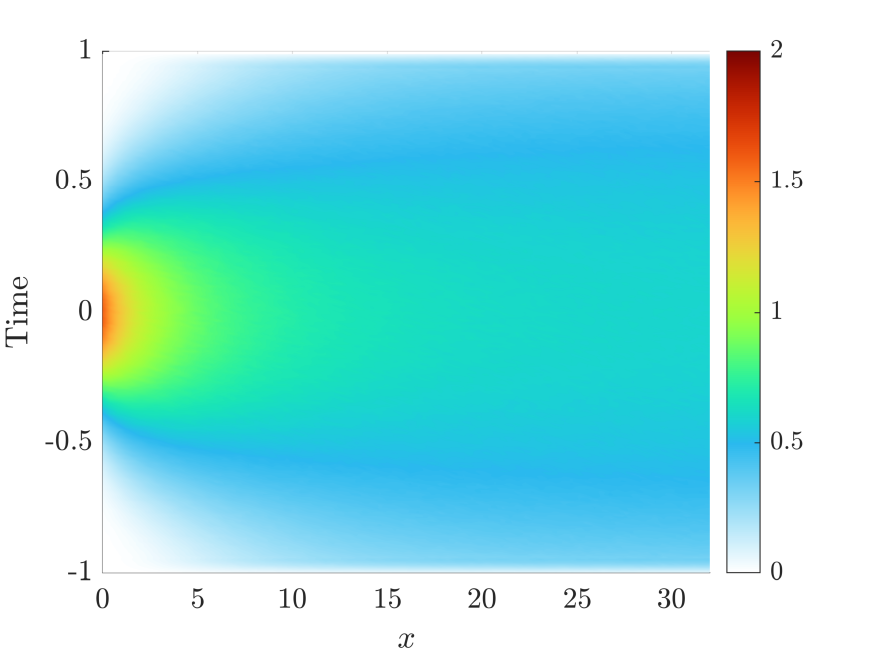}\hfil
\includegraphics[width=0.3\textwidth]{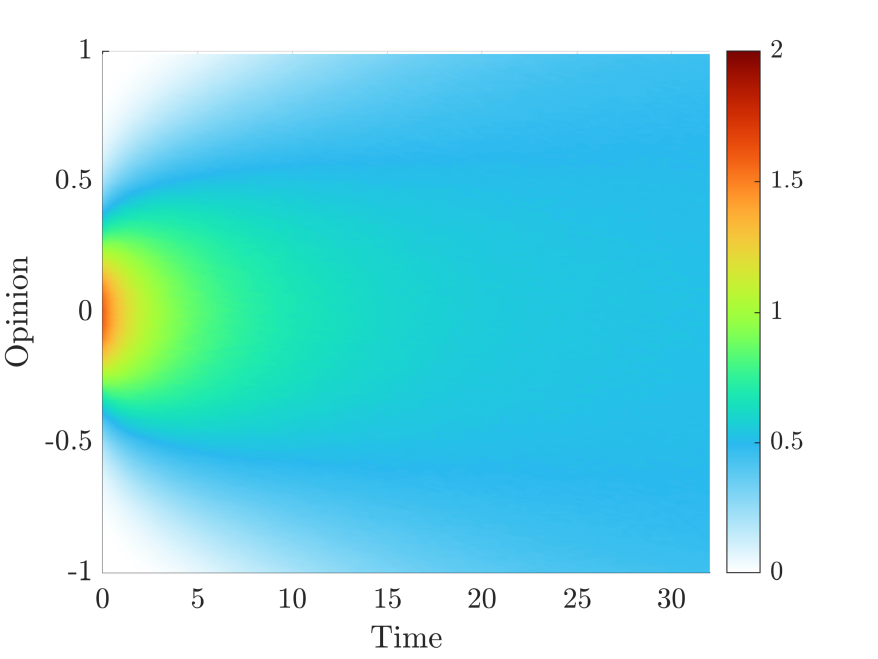}}
\caption{Top row: evolution in time of the distribution function $g(x,w,\tau)$ 
for different time snapshots, respectively $\tau = 0$, $\tau = 16$ and $\tau = 
32$. Bottom row: evolution in time of the opinion marginal $g(w,\tau)$. 
Columns, from left to right: simulations results for $\epsilon = 10^{-1}$, 
$\epsilon = 10^{-2}$ and $\epsilon = 5\times 10^{-4}$.}
\label{fig:pl-evolution}
\end{figure}
In Figure~\ref{fig:pl-entropy} we report again for ease of viewing the 
power-law graphon kernel that we use in our simulations and the time evolution 
of the entropy, computed as
\[
H[g](\tau) = -\intz \f\log \f\, dx\, dw.
\]
Again, we see that in the limit $\epsilon \to 0^+$ the state reaches an uniform 
distribution over $\Omega\times\I = [0, 1]\times[-1, 1]$, since the power-law 
graphon is defined on the entire unit square~$[0, 1]^2$.
\begin{figure}[htbp]
\setbox0=\hbox{\includegraphics[width=0.45\textwidth]{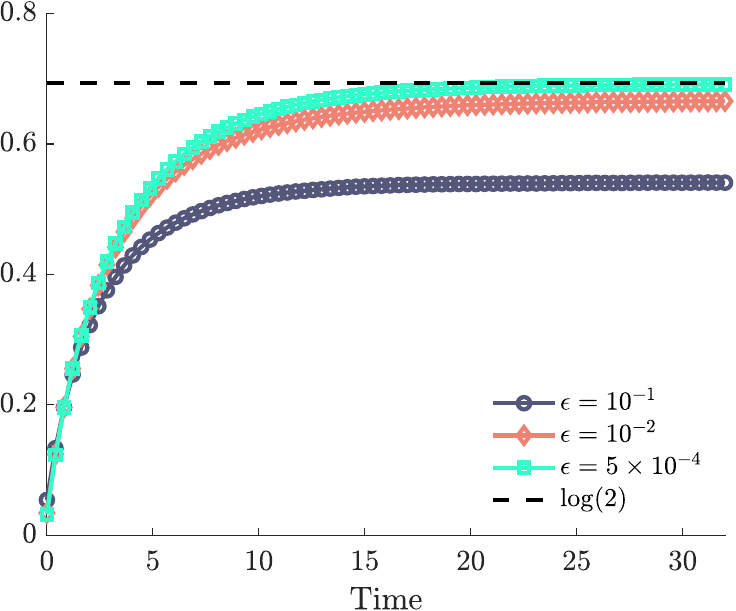}}
\centering
\hbox to \textwidth{%
\includegraphics[width=0.45\textwidth,height=\ht0]{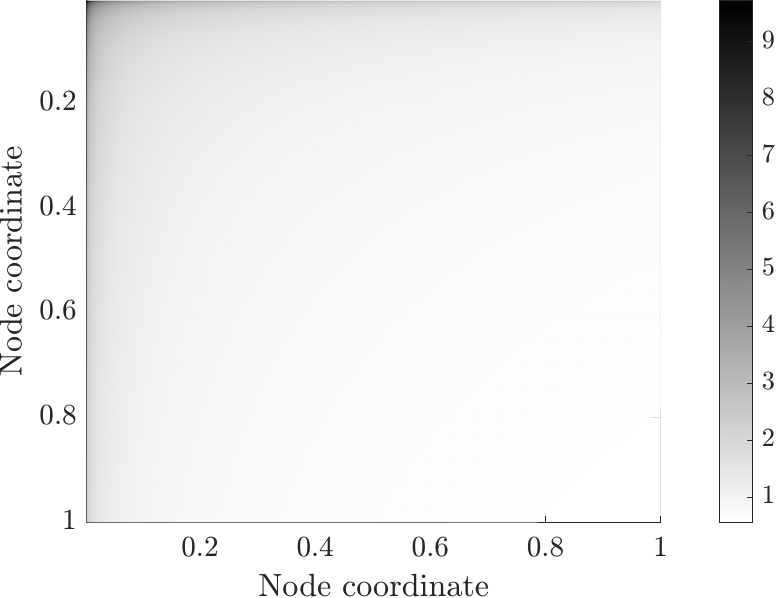}\hfil
\includegraphics[width=0.45\textwidth]{pl-entropy-new-eps-converted-to.pdf}}
\caption{Left: surface plot of $\B(x,y) = 9/16(xy)^{-1/4}$. Right, time 
evolution of the entropy~$H[g](\tau)$ for the same values of~$\epsilon$ of 
Figure~\ref{fig:pl-evolution}.}
\label{fig:pl-entropy}
\end{figure}

\subsection{Declustering on non-separable networks}

The last computational experiments illustrate the dynamics in case of 
non-separable graphons: the \kNN\ graphon and the small-world graphon. The first one  models the \kNN\ networks as 
described, e.g., in \cite{dedios22, watts98, cambridge-book} and it is defined (\cite{dedios22}) as
\[
\mathop\B\limits^{\hbox{\tiny \kNN}}(x,y) = (1-p)\mathop\B\limits^{\hbox{\tiny \itshape SW}}(x,y) + p(1 - \mathop\B\limits^{\hbox{\tiny \itshape SW}}(x,y)),
\]
where $\chi(\phv)$ is the indicator function and $p, r \in (0, 1)$ are constant 
real numbers, while the small-world graphon (see, e.g., \cite{dedios22, watts98, cambridge-book}) can be defined as
\[
\mathop\B\limits^{\hbox{\tiny\itshape SW}}(x,y) = \chi(\min(\abs{x-y}, 1 - 
\abs{x-y}) \le r).
\]
\begin{figure}[htbp]
\setbox0=\hbox{\includegraphics[width=0.3\textwidth]{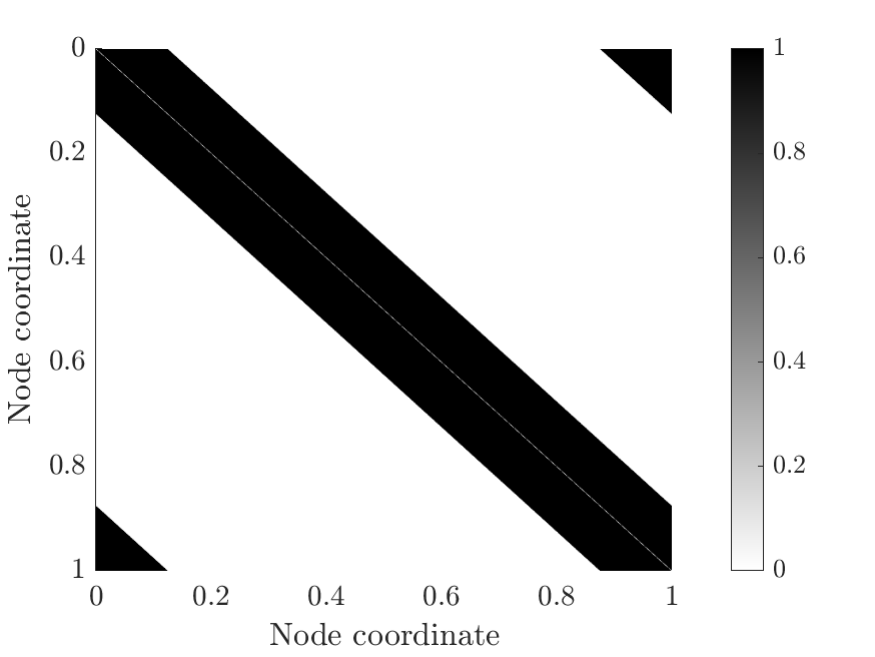}}
\centering
\hbox to \textwidth{%
\includegraphics[width=0.3\textwidth, trim=0 0 1.05cm 0.8cm] 
{small-world-graphon.pdf}\hfil
\includegraphics[width=0.3\textwidth, 
height=\ht0]{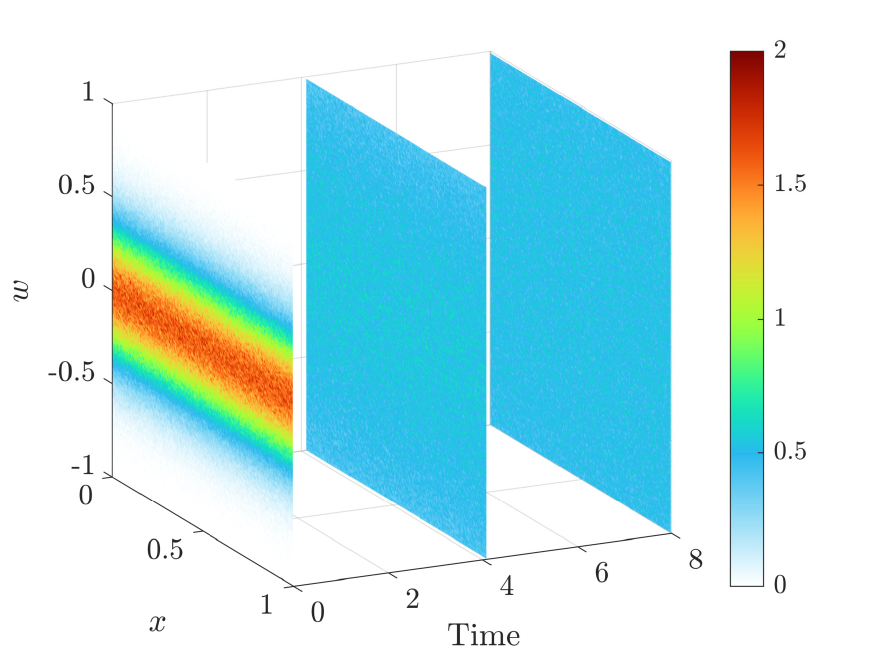}\hfil
\includegraphics[width=0.3\textwidth, 
height=\ht0]{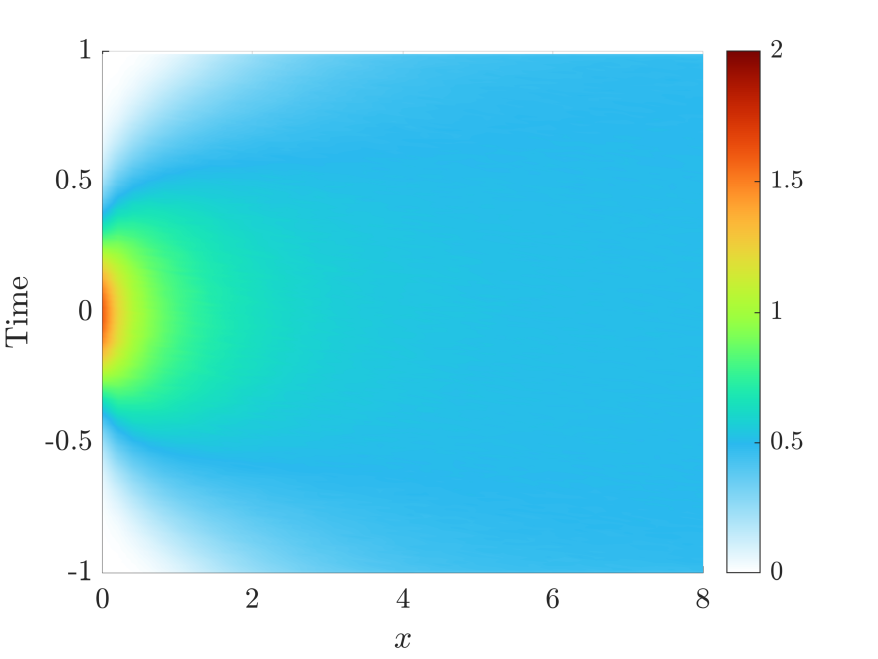}}
\hbox to \textwidth{%
\includegraphics[width=0.3\textwidth, trim=0 0 1.05cm 0.8cm,
]{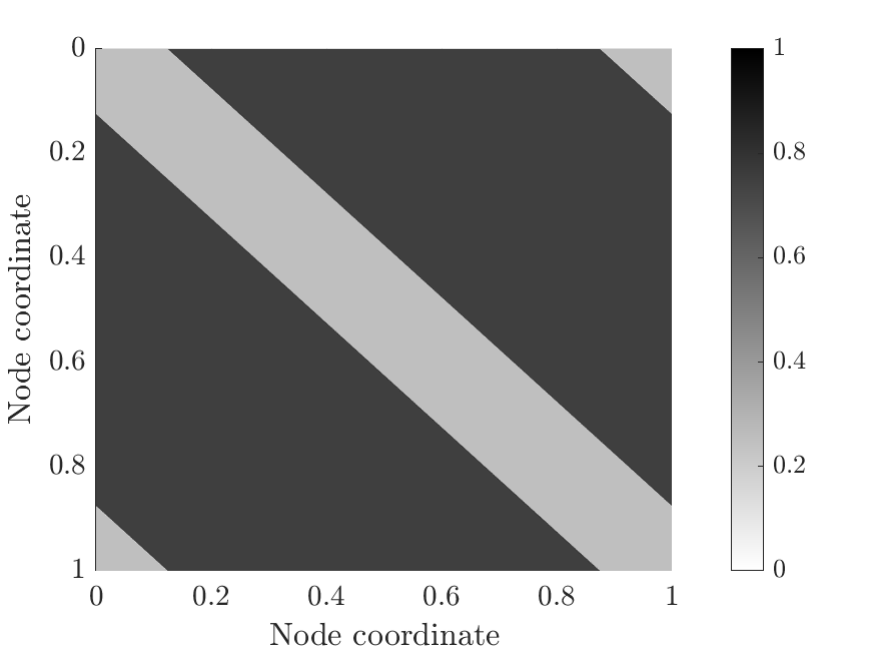}\hfil
\includegraphics[width=0.3\textwidth, 
height=\ht0]{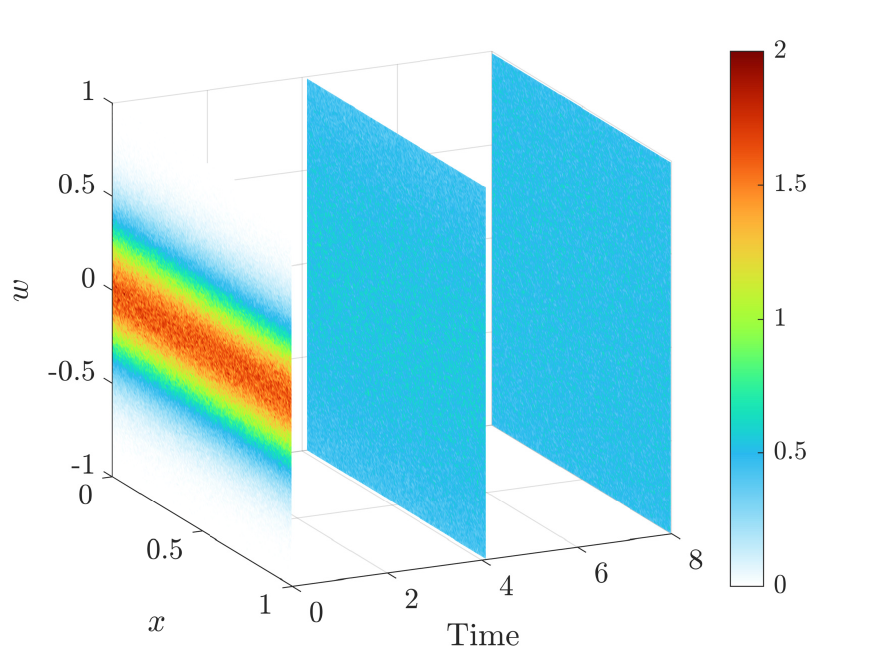}\hfil
\includegraphics[width=0.3\textwidth, 
height=\ht0]{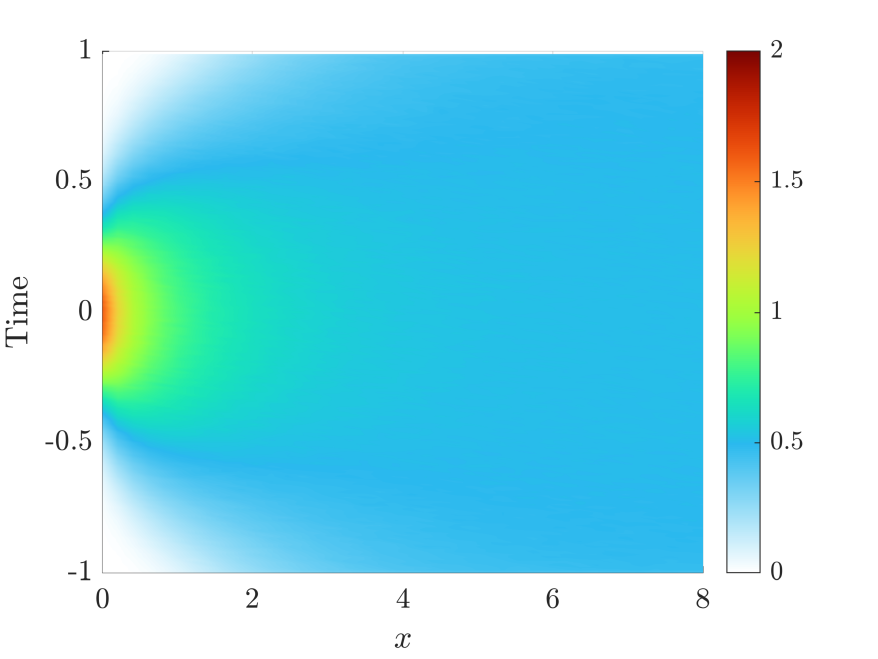}}
\caption{Top row: simulation results for the small-world graphon kernel. 
Bottom row: simulation results for the \kNN\ graphon kernel. Column-wise, 
from left to right: surface plot of the graphon kernel; time snapshots slices 
of the distribution $\f$ for $\tau =0$, $\tau = 4$ and $\tau = 8$; time 
evolution of the opinion marginal distribution $g(w,\tau)$.}
\label{fig:non-separable-evolution}
\end{figure}
\!In Figure~\ref{fig:non-separable-evolution} we reported the surface plots 
for 
both graphon kernels for fixed values of $r = 1/8$ and $p = 3/4$. We consider 
model~\eqref{eq:general-control-FP} with scaling 
parameter~$\epsilon=10^{-3}$ and let evolve in time until~$T = 8$, where 
for this test we considered the network and time dependent optimal penalty 
coefficient~$\bar\kappa_{x,\tau}$ as written in 
equation~\eqref{eq:kappa-bar-general}.
As we can see, $\f$ approaches a uniform distribution over both network 
topologies.

\section*{Conclusion}

In this paper we proposed a simple yet very efficient optimal control strategy 
to break consensus in a kinetic model for opinion formation on graphons. The 
proposed approach allows us to include complex microscopic features, such as 
social networks, on the continuum limit and understand the impact of simple 
declustering mechanisms.

In doing so, we investigate the uncontrolled and controlled models in the 
mean-field limit. We then investigate the large time behavior and are able to 
write down closed form solutions of the (quasi)-stationary agent distribution 
of the uncontrolled and controlled problem for certain choices of parameters. 
This formulation allows us to identify the necessary controls to prevent 
consensus and steer the crowd toward a uniform distribution. We corroborate 
our analytical results with computational experiments for various types of 
graphons. The numerical results confirm our theoretical findings and the 
success of the proposed declustering strategy. Extensions of the designed 
approach to include dynamic networks for fully nonlinear equations are actually 
under study and will be presented in future researches.  

\section*{Acknowledgments}
This work have been written within the activities of the Royal Society 
Grant for International Exchanges Ref.\ IES/R3/213113.
M.Z. and J.F. acknowledge the support of MUR-PRIN2020 Project 
No.~2020JLWP23 and of the GNFM group of INdAM (National Institute of High Mathematics). M.Z. acknowledges the support of Next Generation EU. M.-T.W. acknowledges partial support from the EPSRC Small 
Grant EPSRC EP/X010503/1.

\appendix
\section{Basic definitions on graphs and graphons}\label{sec:graphons}

We follow here~\cite{borgs14,glasscock15} to give a brief overview on graphs 
and graphons.

A (simple, unweighted) graph $G$ is a pair of sets: $V(G)$ that indicates 
vertices, or nodes, of $G$, and $E(G)$ that refers to the edges between nodes, 
pairwise distinct. Two connected vertices are also said to be \emph{adjacent}, 
and in this spirit we can describe the graph $G$ via its associated 
\emph{adjacency matrix} $A_{ij}(G)$, where $A_{ij} = 1$ if and only if $(i,j) 
\in E(g)$, that is, if nodes $i$ and $j$ are connected. A way to represent the 
adjacency matrix of a graph is its \emph{pixel picture}, i.e., we discretize 
the unit square $[0,1]^2 \subset \R$ into a grid of $N^2$ squares of size 
$1/N$, where $N$ is the number of nodes of $G$. Then the square whose 
north-west coordinate is $\bigl((i-1)/N, (j-1)/N\bigr)$ is painted black if and 
only if $A_{ij}(G) = 1$. Since matrices are labeled starting from their 
upper-left corner, the same happens for the pixel pictures, where the origin is 
at their upper-left corner as well.
\begin{figure}[hbp]
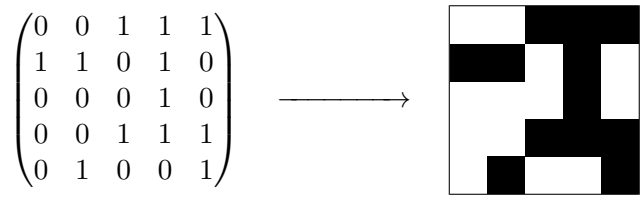

    \centering
    $\displaystyle
    \begin{pmatrix}
    0 & 0 & 1 & 1 & 1\\
    1 & 1 & 0 & 1 & 0\\
    0 & 0 & 0 & 1 & 0\\
    0 & 0 & 1 & 1 & 1\\
    0 & 1 & 0 & 0 & 1\\
    \end{pmatrix}
    \quad\xrightarrow{\rule{4em}{0pt}}\quad
\hbox{\vrule$\vcenter{%
\let\riga\hbox
\parindent=0pt
\topskip=0pt
\parindent=0pt
\offinterlineskip
\catcode`\0=\active
\catcode`\1=\active
\def0{\hspace*{3ex}}
\def1{\rule{3ex}{3ex}}
\hrule
\riga{00111}
\riga{11010}
\riga{00010}
\riga{00111}
\riga{01001}
\hrule}$\vrule}
    $
    \caption{Example of pixel picture of a $5\times5$ adjacency matrix.}
    \label{fig:adjmat1}
\end{figure}

If we let the number $N$ become large, we see that the pixel picture \lq\lq 
converges\rq\rq\ toward a grayscale image. For example, if we consider a random 
graph  where each node is linked to another with probability $1/2$ seems to 
approach the constant function $1/2$ over $[0,1]^2$. This concept of limit of a sequence of graphs can be made rigorous introducing 
the notion of \emph{graphon}.

\begin{figure}
    \centering
    \includegraphics[width=0.9\textwidth]{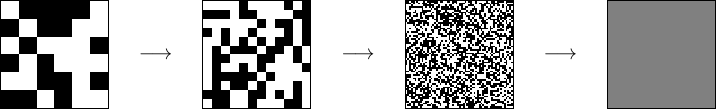}
    \caption{Convergence of pixel pictures of adjacency matrices of random 
    graphs to the function $1/2$.}
    \label{fig:adjmat2}
\end{figure}

We define a (labeled) graphon as a Lebesgue-measurable function from 
$\Omega^2\subseteq [0,1]^2$ to $\R^+$, with the tacit assumption that we 
identify graphons that are equal almost everywhere. An unlabeled graphon is a 
labeled graphon to which is applied an invertible, measure preserving map to 
$[0,1]$ (called \emph{re-labeling}). 

A familiar example of a graphon is just a finite graph, so that it is sensible 
to see graphons as a consistent generalization of graphs. Indeed, finite graphs 
can be described as \emph{stepfunctions} on $[0,1]^2]$, i.e., measurable 
functions that are piecewise constant. We can construct the step 
graphon~$W^G$ associated to the finite graph~$G$ directly, pretty much like 
we did to pass from adjacency matrix to pixel picture:
\(
W^G \coloneqq \frac{1}{\abs{V(G)}^2}\sum_{i,j\in V(G)} \chi_{\strut I_i\times 
I_j},
\label{eq:stepgraph}
\)
where $I_i \subseteq [0,1]$ is an interval of length $1/\abs{V(G)}$.

Graphons themselves become elements of a metric space (which we refer to as 
$(\G,\cut)$) once we consider a distance between pairs $(W,U)$ of them, called 
the \emph{cut distance}
\[
\cut(W,U) = \inf_{\varphi,\psi}\, \sup_{S,T}\,
\abs[\bigg]{\int\limits_{S\times T}\! W\bigl(\varphi(x),\varphi(y)\bigr) - 
U\bigl(\psi(x),\psi(y)\bigr)\, dx\,dy},
\]
where $\varphi$ and $\psi$ are re-labeling while $S$ and $T$ are measurable 
subsets of $\Omega \subseteq [0,1]$. Notice that the cut distance is just a 
pseudo-metric, since $\cut(W,U) = 0 \nimplies  W = U$.

There are two main results that are useful when dealing with 
graphons~\cite{glasscock15}, which we report for convenience.
\begin{thm}
Given a graphon $W$ there exists a sequence $(W_n)_n$ such that
\[
\cut{(W_n - W)} \xrightarrow[n\to+\infty]{} 0.
\]
\end{thm}
\begin{thm}
The space $(\G,\cut)$ is compact. 
\end{thm}
This implies that the space of graphons endowed with the cut metric is 
complete: indeed, it is the completion of the space of finite graphs when 
itself is equipped with the cut norm.

While these results give us a nice framework to analyze graphs when their 
number of vertices grows very large, an issue arises when we consider the limit 
of a sequence of \emph{sparse} graphs. The (edge) density of a graph can be 
defined as the fraction of its number of edges $\abs{E(G)}$ with respect to the 
maximum possible, which for simple, directed graphs is 
$\abs{V(G)}(\abs{V(G)}-1)$
\[
d(G) \coloneqq \frac{\abs{E(G)}}{\abs{V(G)}(\abs{V(G)}-1)}.
\]
A dense graph is one such that for its two magnitudes $\abs{E(G)}$ and 
$\abs{V(G)}$ there is a relation like $\abs{E(G)} = O(\abs{V(G)}^2)$, while for 
a sparse graph holds $\abs{E(G)} = o(\abs{V(G)}^2)$. For sequences of graphs 
$(G_n)_n$ this translates into 
\[
\begin{aligned}
    \text{$(G_n)_n$ is dense} \iff \liminf_{n\to +\infty} d(G_n) > 0; &&
    \text{$(G_n)_n$ is sparse} \iff \lim_{n\to +\infty} d(G_n) = 0 
\end{aligned}
\]
The density of a graph can be rewritten as a sum
\(
d(G) = \sum_{i,j\in V(G)} \frac{1}{\abs{V(G)}(\abs{V(G)}-1)}
\label{eq:density1}
\)
The form~\eqref{eq:density1} of the density of a graph is a particular case 
of the $p$-norm for weighted graphs
\(
\norm{G}_p \coloneqq
\begin{cases}\displaystyle
 \Biggl(\,\sum_{i,j\in V(G)}\frac{\alpha_i\alpha_j}{\alpha_G(\alpha_G - 
 1)}\abs{\beta_{ij}^p}\Biggr)^{\kern-2pt 1/p}, & 1\le p < \infty\\\displaystyle
\max_{i,j\in V(G)} \abs{\beta_{ij}}, & p = \infty,
\end{cases}
\label{eq:pnormG}
\)
where $\alpha_i$ is the weight of the $i$-th node, $\beta_{ij}$ is the weight 
of the edge that connects node~$i$ with node~$j$ (with $\beta_{ij} = 0$ if 
the two are not adjacent) and finally $\alpha_G$ is the total weight of the 
graph
\[
\alpha_G = \sum_{i\in V(G)}\alpha_i.
\]
Now we see that when we consider an unweighted, directed, simple graph, 
equation~\eqref{eq:density1} is equation~\eqref{eq:pnormG} when $p$ equal 
$1$.
We may define the $p$-norm for graphons, too: for $1\le p < +\infty$ it has the 
form
\[
\norm{W}_p \coloneqq \biggl(\ \int_\Omega \abs{W(x,y)}^p\, dx\,dy\biggr)^{1/p},
\]
while $\norm{W}_\infty$ is the essential supremum of $W$.
Notice that $\norm{W^G}_p = \norm{G}_p$ for all $p \in [1,+\infty]$. This tells 
us that a sequence of sparse graphs will converge toward one of vanishing 
density, so that its limit graphon will effectively be the uninteresting null 
graphon.

To overcome this difficulty, we will consider the \emph{normalized} graphs, 
i.e., we will divide them by their $1$-norm: this will allow us to compare 
graphons with different densities and to consider meaningful sequences of 
sparse graphs, with the aim of introducing and studying \emph{sparse} graphons.

Results in \cite{borgs14} assure that normalized graphs converge, up to 
subsequences, to an $L^p$ graphon in the cut metric; so do sequences of step 
graphons. Moreover, the $L^p$ ball of graphons is compact with respect to the 
cut metric (up to identification of objects at zero distance).

At last, we recall the concept of \emph{connectivity} and \emph{connected 
component} 
for a graphon.
\begin{definition}
Given a graphon $W \colon \Omega^2 \subseteq [0,1]^2 \to \R^+$ we say it is 
\emph{disconnected} if there exists a set $S \subset \Omega$ such that $S$ has 
positive Lebesgue measure and
\[
\mathclap{\int\limits_{S \times (\Omega \setminus S)}}\enspace W(x,y) \, dx\,dy 
= 0.
\]
We say that $W$ is \emph{connected} if it is not disconnected. A subset $S 
\subset \Omega$ such that $S$ is connected and $S \cup \{x\}$, for all $x \in 
\Omega \setminus S$, is not, is called \emph{connected component} of the 
graphon $W$.
\end{definition}
Clearly, we can always write a graphon as a disjoint union of its connected 
components.

\begin{definition}\label{def:degree}
Given a graphon $W \colon \Omega^2 \subseteq [0,1]^2 \to  \R^+$ and point 
$x$ 
and $y \in \Omega$, we define 
\[
d_i(x) = \int_\Omega W(x,y)\, dy \qquad d_o(y) = \int_\Omega W(x,y)\, dx
\]
as the \emph{in-degree} of $x$ and the \emph{out-degree} of $y$  with respect 
to the graphon $W(x,y)$, respectively. The two definitions agree in case of 
undirected, i.e., symmetric, graphons.
\end{definition}

%\nocite{*}

%\bibliographystyle{abbrv}
%\bibliography{refs.bib}
\printbibliography[heading=bibintoc, title={References}]

@book{CIP94,
  title={The Mathematical Theory of Dilute Gases},
  author={Cercignani, C. and Illner, R. and Pulvirenti, M.},
  isbn={9780387942940},
  lccn={lc94010086},
  series={Applied Mathematical Sciences},
  url={https://books.google.com/books?id=F5NINMV-9IsC},
  year={1994},
  publisher={Springer New York}
}

@book{Cerc88,
  title={The {B}oltzmann Equation and Its Applications},
  author={Cercignani, C.},
  isbn={9781461210399},
  lccn={87026654},
  series={Applied Mathematical Sciences},
  url={https://books.google.com/books?id=OcTcBwAAQBAJ},
  year={2012},
  publisher={Springer New York}
}

@article{medvedev14,
author = {Medvedev, Georgi S.},
title = {The Nonlinear Heat Equation on Dense Graphs and Graph Limits},
journal = {SIAM Journal on Mathematical Analysis},
volume = {46},
number = {4},
pages = {2743-2766},
year = {2014}}

@article{petit21,
author = {Petit, Julien and Lambiotte, Renaud and Carletti, Timoteo},
title = {Random Walks on Dense Graphs and Graphons},
journal = {SIAM Journal on Applied Mathematics},
volume = {81},
number = {6},
pages = {2323-2345},
year = {2021}}

@article{ACFK,
	author = {Albi, G. and Choi, Y.-P. and Fornasier, M. and Kalise, D.},
	journal = {Appl. Math. Optim.},
	pages = {93--135},
	title = {Mean Field Control Hierarchy},
	volume = {76},
	year = {2017}}

@article{BL,
	author = {Borra, D. and Lorenzi, T.},
	journal = {Commun. Pure Appl. Anal.},
	number = {3},
	pages = {1487--1499},
	title = {Asymptotic analysis of continuous opinion dynamics models under bounded confidence},
	volume = {12},
	year = {2013}}

@article{zanella_BMB23,
	author = {Zanella, M.},
	journal = {Bull. Math. Biol.},
	number = {36},
	title = {Kinetic Models for Opinion Dynamics in the Presence of Opinion Polarization},
	volume = {85},
	year = {2023}}

@article{CPT,
	author = {Cordier, S. and Pareschi, L. and Toscani, G.},
	journal = {J. Stat. Phys.},
	number = {112},
	pages = {253--277},
	title = {On a Kinetic Model for a Simple Market Economy},
	volume = {120},
	year = {2005}}

@article{newman03,
	author = {Newman, Mark E. J.},
	journal = {SIAM Rev.},
	number = {2},
	pages = {167--256},
	title = {The Structure and Function of Complex Networks},
	volume = {45},
	year = {2003}}

@incollection{CFTV,
	author = {Carrillo, J. A. and Fornasier, M. and Toscani, G. and Vecil, F.},
	booktitle = {Mathematical Modeling of Collective Behavior in Socio-Economic and Life Sciences},
	editor = {Naldi, G. and Pareschi, L. and Toscani, G.},
	publisher = {Birkh{\"a}user Boston},
	series = {Modeling and Simulation in Science, Engineering and Technology},
	title = {Particle, kinetic, and hydrodynamic models of swarming},
	year = {2010}}

@article{CFRT,
	author = {Carrillo, J. A. and Fornasier, M. and Rosado, J. and Toscani, G.},
	journal = {SIAM J. Math. Anal.},
	number = {1},
	pages = {218--236},
	title = {Asymptotic flocking dynamics of the kinetic {C}ucker-{S}male model},
	volume = {42},
	year = {2010}}

@article{albi17,
	author = {Albi, Giacomo and Pareschi, Lorenzo and Zanella, Mattia},
	journal = {Kinet. Relat. Mod.},
	number = {1},
	pages = {1--32},
	title = {Opinion dynamics over complex networks: kinetic modelling and numerical methods},
	volume = {10},
	year = {2017}}

@article{TTZ18,
	author = {Toscani, Giuseppe and Tosin, Andrea and Zanella, Mattia},
	journal = {Phys. Rev. E},
	number = {2},
	pages = {022315},
	title = {Opinion modeling on social media and marketing aspects},
	volume = {98},
	year = {2018}}

@inproceedings{APZ16,
	author = {Albi, Giacomo and Pareschi, Lorenzo and Zanella, Mattia},
	booktitle = {System Modeling and Optimization. CSMO 2015. IFIP Advances in Information and Communication Technology},
	editor = {Bociu, L. and D{\'e}sid{\'e}ri, J.A. and Habbal, A.},
	title = {On the optimal control of opinion dynamics on evolving networks},
	volume = {494},
	year = {2016}}

@article{GGSP,
	author = {Goddard, B. D. and Gooding, B. and Short, H. and Pavliotis, G. A.},
	journal = {IMA J. Appl. Math.},
	number = {1},
	pages = {80--110},
	title = {Noisy bounded confidence models for opinion dynamics: the effect of boundary conditions on phase transitions},
	volume = {87},
	year = {2021}}

@article{APZ14,
	author = {Albi, Giacomo and Pareschi, Lorenzo and Zanella, Mattia},
	journal = {Phil. Trans. R. Soc. A},
	number = {2028},
	title = {Boltzmann-type control of opinion consensus through leaders},
	volume = {372},
	year = {2014}}

@article{glasscock15,
	author = {Glasscock, Daniel},
	journal = {Notices of the AMS},
	number = {1},
	pages = {46--48},
	title = {What is\ldots\ a graphon},
	volume = {62},
	year = {2015}}

@article{borgs14,
	author = {Christian Borgs and Jennifer T. Chayes and Henry Cohn and Yufei Zhao},
	journal = {Trans. Am. Math. Soc.},
	title = {An $L^p$ theory of sparse graph convergence I: Limits, sparse random graph models, and power law distributions},
	year = {2014}}

@article{toscani06,
	author = {Toscani, Giuseppe},
	journal = {Commun. Math. Sci.},
	number = {3},
	pages = {481--496},
	publisher = {International Press of Boston},
	title = {Kinetic models of opinion formation},
	volume = {4},
	year = {2006}}

@article{bonnet22,
	author = {Bonnet, Beno{\^\i}t and Pouradier Duteil, Nastassia and Sigalotti, Mario},
	journal = {Math. Mod. Meth. Appl. Sci.},
	number = {11},
	pages = {2121--2188},
	publisher = {World Scientific},
	title = {Consensus formation in first-order graphon models with time-varying topologies},
	volume = {32},
	year = {2022}}

@article{piccoli19,
	author = {Piccoli, Benedetto and Pouradier Duteil, Nastassia and Tr{\'e}lat, Emmanuel},
	journal = {SIAM J. Contr. Optim.},
	number = {4},
	pages = {2628--2659},
	publisher = {SIAM},
	title = {Sparse control of {H}egselmann-{K}rause models: Black hole and declustering},
	volume = {57},
	year = {2019}}

@book{pareschi13,
	author = {Pareschi, Lorenzo and Toscani, Giuseppe},
	publisher = {OUP Oxford},
	title = {Interacting Multiagent Systems: Kinetic Equations and Monte Carlo Methods},
	year = {2013}}

@unpublished{dedios22,
	author = {de Dios, Blanca Ayuso and Dovetta, Simone and Spinolo, Laura V},
	note = {Preprint arXiv:2211.01932},
	title = {On the continuum limit of epidemiological models on graphs: convergence results, approximation and numerical simulations},
	year = {2022}}

@article{bondesan24,
	author = {Bondesan, Andrea and Toscani, Giuseppe and Zanella, Mattia},
	journal = {Math. Mod. Meth. Appl. Sci.},
	publisher = {World Scientific},
	title = {Kinetic compartmental models driven by opinion dynamics: vaccine hesitancy and social influence},
	year = {2024}}

@article{deffuant04,
	author = {Fr{\'e}d{\'e}ric Amblard and Guillaume Deffuant},
	issn = {0378-4371},
	journal = {Phys. A},
	pages = {725-738},
	title = {The role of network topology on extremism propagation with the relative agreement opinion dynamics},
	volume = {343},
	year = {2004}}

@article{amelkin17,
	author = {Amelkin, Victor and Bullo, Francesco and Singh, Ambuj K},
	journal = {IEEE Trans. Automat. Contr.},
	number = {11},
	pages = {5650--5665},
	publisher = {IEEE},
	title = {Polar opinion dynamics in social networks},
	volume = {62},
	year = {2017}}

@article{degroot74,
	author = {DeGroot, Morris H},
	journal = {J. Amer. Stat. Assoc.},
	number = {345},
	pages = {118--121},
	publisher = {Taylor \& Francis},
	title = {Reaching a consensus},
	volume = {69},
	year = {1974}}

@article{lee14,
	author = {Lee, Jae Kook and Choi, Jihyang and Kim, Cheonsoo and Kim, Yonghwan},
	journal = {J. Commun.},
	number = {4},
	pages = {702--722},
	publisher = {Oxford University Press},
	title = {Social media, network heterogeneity, and opinion polarization},
	volume = {64},
	year = {2014}}

@article{matakos17,
	author = {Matakos, Antonis and Terzi, Evimaria and Tsaparas, Panayiotis},
	journal = {Data Mining and Knowledge Discovery},
	pages = {1480--1505},
	publisher = {Springer},
	title = {Measuring and moderating opinion polarization in social networks},
	volume = {31},
	year = {2017}}

@article{sznajd00,
	author = {Sznajd-Weron, Katarzyna and Sznajd, Jozef},
	journal = {Int. J. Mod. Phys. C},
	number = {06},
	pages = {1157--1165},
	publisher = {World Scientific},
	title = {Opinion evolution in closed community},
	volume = {11},
	year = {2000}}

@article{hu23,
	author = {Hu, Yuanquan and Wei, Xiaoli and Yan, Junji and Zhang, Hengxi},
	journal = {J. Frank. Inst.},
	number = {18},
	pages = {14783--14805},
	publisher = {Elsevier},
	title = {Graphon mean-field control for cooperative multi-agent reinforcement learning},
	volume = {360},
	year = {2023}}

@article{coppini22,
	author = {Coppini, Fabio},
	journal = {J. Stat. Phys.},
	number = {2},
	pages = {15},
	publisher = {Springer},
	title = {A Note on {F}okker-{P}lanck Equations and Graphons},
	volume = {187},
	year = {2022}}

@article{gao19,
	author = {Gao, Shuang and Caines, Peter E},
	journal = {IEEE Trans. Automat. Contr.},
	number = {10},
	pages = {4090--4105},
	publisher = {IEEE},
	title = {Graphon control of large-scale networks of linear systems},
	volume = {65},
	year = {2019}}

@article{erol23,
	author = {Erol, Selman and Parise, Francesca and Teytelboym, Alexander},
	issue = {105673},
	journal = {J. Econ. Theory},
	publisher = {Elsevier},
	title = {Contagion in graphons},
	year = {2023}}

@article{bayraktar23,
	author = {Bayraktar, Erhan and Chakraborty, Suman and Wu, Ruoyu},
	journal = {Ann. Appl. Probab.},
	number = {5},
	pages = {3587--3619},
	publisher = {Institute of Mathematical Statistics},
	title = {Graphon mean field systems},
	volume = {33},
	year = {2023}}

@inproceedings{alothali18,
	author = {Alothali, Eiman and Zaki, Nazar and Mohamed, Elfadil A and Alashwal, Hany},
	booktitle = {2018 International Conference on Innovations in Information Technology (IIT)},
	organization = {IEEE},
	pages = {175--180},
	title = {Detecting social bots on twitter: a literature review},
	year = {2018}}

@article{gilani19,
	author = {Gilani, Zafar and Farahbakhsh, Reza and Tyson, Gareth and Crowcroft, Jon},
	journal = {ACM Transactions on the Web (TWEB)},
	number = {1},
	pages = {1--23},
	publisher = {ACM New York, NY, USA},
	title = {A large-scale behavioural analysis of bots and humans on twitter},
	volume = {13},
	year = {2019}}

@article{motsch14,
	author = {Motsch, Sebastien and Tadmor, Eitan},
	journal = {SIAM Rev.},
	number = {4},
	pages = {577--621},
	publisher = {SIAM},
	title = {Heterophilious dynamics enhances consensus},
	volume = {56},
	year = {2014}}

@article{caron23,
	author = {Caron, Fran{\c{c}}ois and Panero, Francesca and Rousseau, Judith},
	journal = {Adv. Appl. Probab.},
	number = {4},
	pages = {1211--1253},
	publisher = {Cambridge University Press},
	title = {On sparsity, power-law, and clustering properties of graphex processes},
	volume = {55},
	year = {2023}}

@book{lovasz12,
	author = {Lov{\'a}sz, L{\'a}szl{\'o}},
	publisher = {American Mathematical Society},
	title = {Large networks and graph limits},
	volume = {60},
	year = {2012}}

@unpublished{naldi22,
	author = {Naldi, Giovanni and Patane, Giuseppe},
	note = {Preprint arXiv:2208.07559},
	title = {A graph-based modelling of epidemics: Properties, simulation, and continuum limit},
	year = {2022}}

@article{pareschi17,
	author = {Pareschi, Lorenzo and Vellucci, Pierluigi and Zanella, Mattia},
	journal = {Phys. A},
	pages = {201--217},
	publisher = {Elsevier},
	title = {Kinetic models of collective decision-making in the presence of equality bias},
	volume = {467},
	year = {2017}}

@article{barabasi99,
	author = {Barab{\'a}si, Albert-L{\'a}szl{\'o} and Albert, R{\'e}ka},
	journal = {Science},
	number = {5439},
	pages = {509--512},
	publisher = {American Association for the Advancement of Science},
	title = {Emergence of scaling in random networks},
	volume = {286},
	year = {1999}}

@article{barabasi09,
	author = {Barab{\'a}si, Albert-L{\'a}szl{\'o}},
	journal = {Science},
	number = {5939},
	pages = {412--413},
	publisher = {American Association for the Advancement of Science},
	title = {Scale-free networks: a decade and beyond},
	volume = {325},
	year = {2009}}

@book{cambridge-book,
	author = {Van Der Hofstad, R.},
	publisher = {Cambridge University Press},
	title = {Random Graphs and Complex Networks},
	volume = {43},
	year = {2016}}

@article{nurisso23,
	author = {Nurisso, Marco and Raviola, Matteo and Tosin, Andrea},
	journal = {Eur. J. Appl. Math.},
	pages = {1--22},
	publisher = {Cambridge University Press},
	title = {Network-based kinetic models: Emergence of a statistical description of the graph topology},
	year = {2024}}

@article{watts98,
	author = {Watts, Duncan J and Strogatz, Steven H},
	journal = {Nature},
	number = {6684},
	pages = {440--442},
	publisher = {Nature Publishing Group},
	title = {Collective dynamics of `small-world'networks},
	volume = {393},
	year = {1998}}

@article{hegselmann02,
	author = {Hegselmann, Rainer and Krause, Ulrich},
	journal = {J. Artif. Soc. Soc. Simul.},
	number = {3},
	title = {Opinion Dynamics and Bounded Confidence Models, Analysis, and Simulation},
	volume = {5},
	year = {2002}}

@article{during09,
	author = {D{\"u}ring, Bertram and Markowich, Peter and Pietschmann, Jan-Frederik and Wolfram, Marie-Therese},
	journal = {Proc. R. Soc. A},
	number = {2112},
	pages = {3687--3708},
	publisher = {The Royal Society Publishing},
	title = {Boltzmann and {F}okker-{P}lanck equations modelling opinion formation in the presence of strong leaders},
	volume = {465},
	year = {2009}}

@article{during15,
	author = {D{\"u}ring, Bertram and Wolfram, Marie-Therese},
	journal = {Proc. R. Soc. A},
	number = {2182},
	pages = {20150345},
	publisher = {The Royal Society Publishing},
	title = {Opinion dynamics: inhomogeneous {B}oltzmann-type equations modelling opinion leadership and political segregation},
	volume = {471},
	year = {2015}}

@article{pareschi19,
	author = {Pareschi, Lorenzo and Toscani, Giuseppe and Tosin, Andrea and Zanella, Mattia},
	journal = {J. Nonlin. Sci.},
	pages = {2761--2796},
	publisher = {Springer},
	title = {Hydrodynamic models of preference formation in multi-agent societies},
	volume = {29},
	year = {2019}}

@article{franceschi231,
	author = {Franceschi, Jonathan AND Pareschi, Lorenzo AND Bellodi, Elena AND Gavanelli, Marco AND Bresadola, Marco},
	journal = {PLoS ONE},
	month = {10},
	number = {10},
	pages = {1-26},
	publisher = {Public Library of Science},
	title = {Modeling opinion polarization on social media: Application to Covid-19 vaccination hesitancy in Italy},
	volume = {18},
	year = {2023}}

@article{albi23,
	author = {Albi, Giacomo and Calzola, Elisa and Dimarco, Giacomo},
	journal = {Eur. J. Appl. Math.},
	pages = {1--27},
	title = {A data-driven kinetic model for opinion dynamics with social network contacts},
	year = {2024}}

@article{iacomini23,
	author = {E. Iacomini and P. Vellucci},
	journal = {J. Math. Sociol.},
	number = {2},
	pages = {123-169},
	publisher = {Routledge},
	title = {Contrarian effect in opinion forming: Insights from {G}reta {T}hunberg phenomenon},
	volume = {47},
	year = {2023}}

@article{albi15,
	author = {Albi, Giacomo and Herty, Michael and Pareschi, Lorenzo},
	journal = {Commun. Math. Sci.},
	number = {6},
	pages = {1407--1429},
	publisher = {International Press of Boston},
	title = {Kinetic description of optimal control problems and applications to opinion consensus},
	volume = {13},
	year = {2015}}

@article{barre18,
	author = {Barr{\'e}, Julien and Carrillo, Jos{\'e} A and Degond, Pierre and Peurichard, Diane and Zatorska, Ewelina},
	journal = {J. Nonlin. Sci.},
	pages = {235--268},
	publisher = {Springer},
	title = {Particle interactions mediated by dynamical networks: assessment of macroscopic descriptions},
	volume = {28},
	year = {2018}}

@article{barre17,
	author = {Barr{\'e}, Julien and Degond, Pierre and Zatorska, Ewelina},
	journal = {Multiscale Model. Simul.},
	number = {3},
	pages = {1294--1323},
	publisher = {SIAM},
	title = {Kinetic theory of particle interactions mediated by dynamical networks},
	volume = {15},
	year = {2017}}

@article{galam08,
	author = {Galam, Serge},
	journal = {Int. J. Mod. Phys. C},
	number = {03},
	pages = {409--440},
	publisher = {World Scientific},
	title = {Sociophysics: A review of {G}alam models},
	volume = {19},
	year = {2008}}

@Article{jabin14,
  author  = {Pierre-Emmanuel Jabin and Sebastien Motsch},
  journal = {J. Diff. Equ.},
  title   = {Clustering and asymptotic behavior in opinion formation},
  year    = {2014},
  issn    = {0022-0396},
  number  = {11},
  pages   = {4165-4187},
  volume  = {257},
}

\end{document}